\documentclass[aps,prd,superscriptaddress,nofootinbib,amsmath,amsfonts,preprintnumbers,groupedaddress,showpacs,10pt,english]{revtex4-1}
\usepackage{amsmath}
\usepackage{amssymb}
\usepackage{babel}
\usepackage{wrapfig}
\usepackage{cancel}

\makeatletter

\usepackage{array,multirow,graphicx}
\usepackage{dcolumn}
\usepackage{newlfont}
\usepackage{bm}
\usepackage[colorlinks,citecolor=blue,urlcolor=blue,linkcolor=blue]{hyperref}
\usepackage[figtopcap]{subfigure}
\usepackage{color}

\begin{document}

%\date{\today}

\title{An innovative black hole solution and thermodynamic properties in higher-order curvature gravity with a scalar field
%An innovative black hole solution in ghost-free $\mathit{f(R,G)}$ gravity  enhanced by scalar field  and Barrow entropy
}

\author{G.G.L. Nashed}
\email{nashed@bue.edu.eg}
\affiliation {Centre for Theoretical Physics, The British University, P.O. Box
43, El Sherouk City, Cairo 11837, Egypt}
\author{Usman Zafar}
\email{zafarusman494@gmail.com,~s2471001@ipc.fukushima-u.ac.jp}
\affiliation{Faculty of Symbiotic Systems Science, Fukushima University,
Fukushima 960-1296, Japan.}
\author{Kazuharu Bamba}
\email{bamba@sss.fukushima-u.ac.jp}
\affiliation{Faculty of Symbiotic Systems Science, Fukushima University,
Fukushima 960-1296, Japan.}
\begin{abstract}
A spherically symmetric black hole solution defined by the gravitational mass is explored in higher-order curvature gravity associated with a scalar field. It is demonstrated that the singularities of the curvature invariants will be far weaker around the central region of the black hole than those in general relativity, owing to the effect originating from higher-order curvatures with the coupling to a dynamical scalar field. Furthermore, thermodynamic properties and the Davies-type phase transition of the black hole are investigated for Barrow entropy with a quantum effect of gravitation. It is found that both the quasi-local energy and Gibbs free energy are positive, and the black hole can hence be stable on the smooth horizon surface. In addition, by analyzing the geodesic deviation, the stability conditions of the black hole are explicitly shown.
%
%Recently, it has been shown that the theory of general relativity (GR) has no success in the confrontations of recent observations. The theory of $\mathit{f(R,G)}$ gravity, where  $\mathit{R}$  and $\mathit{G}$  are the Ricci and Gauss-Bonnet scalars, respectively, is regarded to be one of good candidates that can cure the anomalies that appeared in the conventional GR. In this realm, we apply the equation of motions of $\mathit{f(R,G)}$ gravity to a spherically symmetric spacetime with two unknown functions and derived all relevant physical quantities of this theory without assuming any constraints on the Ricci scalar or on $\mathit{f(R,G)}$ gravity. The black hole of this solution is characterized by the gravitational mass of the system and depends on a dimensional parameter, which makes it deviate from the Schwarzschild geometry of  GR. We show that this extra parameter makes the singularities of the invariants of this theory, $\mathit{f(R, G)}$,   much weaker compared to the black holes of Einstein theory.  We examine this black hole using the framework of traditional thermodynamics procedure as well as the standard Barrow one, showing their consistency with well-known results presented in the literature. We also discuss the thermodynamic topology and derive the topological current by using Duan's theory. Moreover, by using the topological charge,  we show that our black hole is stable in the smooth horizon surface.  Finally, using the geodesic deviations, we derive the stability conditions of this black hole.
%
\end{abstract}

%\pacs{04.50.Kd, 04.25.Nx, 04.40.Nr}

%\begin{center}
\maketitle
\section{\bf Introduction}
%\end{center}  2409.15606v1.

The formulation of general relativity (GR) revolutionized our understanding of gravity, interpreting it as the curvature of spacetime. Despite its success in explaining numerous astrophysical and cosmological phenomena, GR encounters challenges in addressing issues such as dark energy, dark matter, and quantum gravity \cite{Penrose:1964wq,Hawking:1973uf,Joshi:2008zz,LIGOScientific:2017vwq,EventHorizonTelescope:2019dse,EventHorizonTelescope:2022wkp}. These limitations have led to the development of modified theories of gravity, which extend or reformulate GR to explore new possibilities to describe gravitational phenomena \cite{DeFelice:2010aj,Sotiriou:2008rp,Nojiri:2006ri,Cai:2015emx}.

Different gravitational theories have been developed as a result of numerous changes to Einstein's GR theory.  For instance, modifications of the GR action include $f(R)$ gravity \cite{DeFelice:2010aj,Nojiri:2010wj,Capozziello:2011et,Nashed:2014sea,Nojiri:2017ncd,Capozziello:2010zz}, with $R$ being the scalar curvature; $f(T)$ gravity \cite{Cai:2015emx,ElHanafy:2014efn,Awad:2017yod,Nashed:2001cp,Awad:2017ign,Zubair:2015cpa,Nashed:2015pga}, with $T$  being the torsion scalar; $f(R,{\cal T})$ theory, in which ${\cal T}$ is the trace  of matter \cite{Harko:2011kv,Zubair:2015gsb}; and $f(T,{\cal T})$ theory \cite{Saleem:2020acy}, $f(G)$ theory, where $G$ denotes the Gauss-Bonnet scalar  \cite{Cognola:2006eg} among others.
 Exploring the issues of  GR, including the rapid expansion of the universe, the flat rotation curves seen in galaxies, the perplexing pattern of wormholes, and other unexplained occurrences near black holes (BHs), has sparked substantial interest in a variety of modified theories. Scientists have intensely been examined these alternative theories to more accurately describe the universe and its mysteries  \cite{DeFelice:2010aj,Capozziello:2011et,Nojiri:2006ri,universe1020123}. Starobinsky was the first scientist to employ a non-linear form of the Ricci scalar, quadratic form, \cite{Starobinsky:1979ty}.  It was demonstrated that the problem of massive neutron stars can be resolved by $f(R)$ gravity \cite{Astashenok_2013,PhysRevD.89.103509,Astashenok_2015,Astashenok_2017,Nashed:2019yto,Astashenok:2016epm}. The $f(R)$ theory is characterized by an arbitrary function with the Ricci scalar as its first order is well known.  In contrast to GR, the field equations of $f(R)$ gravity offer significant classes of solutions.  The analysis of how matter fields and dark energy evolve and interact within the framework of $f(R)$ gravity has been a significant focus  \cite{Stabile:2013eha,Nashed:2021sji,Nojiri:2006ri}. From a cosmological perspective, numerous researchers have conducted extensive studies and investigations. Their work centers on understanding the fundamental processes and phenomena that shape the universe, including its origins, evolution, and eventual fate. This research is vital for refining our understanding of cosmic events and the laws governing them \cite{Shah:2019mxn,Nojiri:2019dqc,Odintsov:2019ofr,PhysRevD.99.064049,Nascimento:2018sir,Miranda:2018jhu,Astashenok:2018bol,PhysRevD.99.063506,Elizalde:2018now,PhysRevD.99.064025,PhysRevD.99.104046,
Bombacigno:2018tyw,Capozziello:2018ddp,Samanta:2019tjb}. The theory of $f(R,G)$ gravity is one of the well-known extensions of GR that offers an appealing framework \cite{Millano:2023gkt,Nashed:2023ewg,Ilyas:2023rde,deHaro:2023lbq,Capozziello:2023vvr,Nojiri:2024nlx,Nojiri:2021mxf}. Higher-order curvature effects are captured by the addition of the Gauss-Bonnet term, a topological invariant in four dimensions, which makes it especially pertinent in high-energy regimes like those around BHs or in the early cosmos.  Furthermore, linking $f(R,G)$ gravity with a scalar field adds more degrees of freedom, enhancing the dynamics of the theory and making it possible to simulate severe gravitational settings more precisely.
  From string theory to inflationary cosmology, scalar fields are essential to contemporary theoretical physics and frequently appear as low-energy expressions of higher-dimensional processes \cite{Capozziello:2011et,DeFelice:2010aj,Nojiri:2006ri}.

A vacuum BH solution with spherical symmetry in  $f(R)$ theory has been obtained in Refs.~\cite{PhysRevD.74.064022,2018EPJP..133...18N,2018IJMPD..2750074N,Nashed:2018piz}. Capozziello et al. utilized the Noether symmetry approach to derive solutions with spherical symmetry within the framework of $f(R)$ theory \cite{Capozziello:2007ec,2012GReGr..44.1881C}. Axially symmetric vacuum BH solutions have been produced using  Noether symmetry \cite{Capozziello:2011et}. For a particular class of $f(R)$ gravity, non-trivial spherically symmetric BH solutions have also been found \cite{Elizalde:2020icc,Nashed:2019yto,Nashed:2019tuk}. The potential importance of strong gravitational fields in local astrophysical objects is highlighted by the inclusion of $f(R)$ gravity. In this context, a great deal of research has been done on spherically symmetric, static BHs  \cite{Sultana:2018fkw,Canate:2017bao,Yu:2017uyd,Canate:2015dda,Kehagias:2015ata,PhysRevD.82.104026,delaCruzDombriz:2009et} and neutron stars \cite{Feng:2017hje,Resco:2016upv,Capozziello:2015yza,Staykov:2018hhc,Doneva:2016xmf,Yazadjiev:2016pcb,Yazadjiev:2015zia,Yazadjiev:2014cza, Ganguly:2013taa,Astashenok:2013vza,Orellana:2013gn,Arapoglu:2010rz,Cooney:2009rr} using non-linear model of $f(R)$ theory. Interestingly, $f(R)$ theory has a scalar potential of gravitational origin \cite{Chiba:2003ir,PhysRevLett.29.137,Chakraborty:2016gpg,Chakraborty:2016ydo}, and is mathematically equivalent to the Brans-Dicke theories \cite{Brans:1961sx}.

In this study, we explore the implications of $f(R,G)$ gravity coupled with a scalar field for BH physics. BHs, being solutions to Einstein's field equations, serve as natural laboratories for testing modifications to GR under extreme conditions. We derive a new BH solution within this modified framework, incorporating the interplay between the curvature invariants and the scalar field. This solution extends existing BH models, providing insights into how higher-order curvature corrections and scalar field dynamics influence the spacetime geometry \cite{ Kanti:1995vq, Nojiri:2005jg}.

According to classical GR, Bekenstein-Hawking entropy is directly related to the area of the event horizon, based on the assumption of a smooth and continuous spacetime. This framework, however, does not account for possible quantum gravitational influences that could alter the BH's horizon at microscopic levels \cite{Barrow:2020tzx}. In addition, our study is based on $f(R,G)$ gravity, a theory that naturally incorporates higher-order curvature corrections and scalar interactions, making it necessary to extend beyond the traditional Bekenstein-Hawking entropy. These scenarios involving modified gravity highlight the growing significance of quantum gravitational effects and changes to the BH horizon's structure. Barrow entropy provides a phenomenological framework for incorporating such effects through a deformation parameter $\delta$, which modifies the standard entropy-area relation to reflect the fractal properties anticipated at Planckian scales \cite{Barrow:2020tzx,Capozziello:2025axh}. Additionally, we investigate the Davies-type phase transition, characterized by divergences in the heat capacity, which indicates a second-order thermodynamic phase transition \cite{Davies:1989ey}. In classical general relativity using Bekenstein-Hawking entropy, continuous phase transitions like those discussed by Davies are generally not observed. It was first established by Davies, in his analysis of Reissnern-Nordstr\"om and Kerr black holes, that these features arise only under specific conditions involving charge or rotation \cite{Davies:1977bgr} (for more details regarding various types of phase transition and Davis-type phase transition, read these Refs.~\cite{Davies:1989ey,Davies:1977bgr,Bhattacharya:2024bjp,Hawking:1982dh,Lousto:1994jd,Muniain:1995ih,Pavon:1988in,Pavon:1991kh,
Cai:1996df,Cai:1998ep,Bhattacharya:2019awq,Dolan:2011jm,Kubiznak:2012wp,Bhattacharya:2017nru}). Conversely, within our framework based on Barrow entropy and higher-order curvature corrections, second-order transitions emerge more generically, with the deformation parameter playing a critical role, highlighting a richer thermodynamic behavior. Our approach is particularly effective for detecting smooth and continuous changes in BH stability, in contrast to first-order phase transitions characterized by discontinuities or latent heat, which do not naturally emerge in our solution. Consequently, both Barrow entropy and Davies-style transitions contribute to a more detailed perspective on the thermodynamic characteristics of black holes in modified gravitational theories.

Our analysis includes deriving the field equations from the modified action and solving them under spherically symmetric assumptions. The newly formed BH's physical characteristics are examined by analyzing its stability criteria, horizon structure, and thermodynamic behavior. This work contributes to the growing body of research on modified gravity theories and offers a platform for testing their predictions in the strong-field regime. This study has the following structure: For $f(R,G)$ gravity, we provide the basic principles in Sec.~\ref{S2}. To study $f(R,G)$ gravity, we have applied its field equations to a spherically symmetric line-element Sec .~\ref {S3}.  Three differential equations in five unknown functions make up the system of differential equations that we derive.  To close the system and obtain the solution, we assume two additional restrictions.
%We go back to the Schwarzschild solution, the BH of GR, if this parameter disappears.  The influence of higher-order curvature, which defines $f(R,G)$ gravity, is represented by this parameter.
In Sec.~\ref{iv}, we compute all the invariants of GR and also the Gauss-Bonnet invariant. When compared to the BHs of GR, we demonstrate that these invariants have stronger singularity as $r \to \infty$.  We compute thermodynamic values such as entropy, Hawking temperature, heat capacity, etc., in Sec.~\ref{S4} and demonstrate that they all behave physically.  Additionally, we compute the topological charge and give  Barrow entropy analysis, proving that our BH is stable in the smooth horizon surface. We have established the criteria for stability for the BHs derived from the $f(R, G)$ gravity theory in Sec.~\ref{S66} using the geodesic deviation. By analyzing these conditions, we gain a better understanding of the dynamics and characteristics of BHs under this modified gravitational framework. Our final thoughts are provided in the final Sec.~\ref{S77}. In Appendix~{\color{blue}A}, for the sake of readability in the main text, we provide detailed explanations of longer equations and the procedures for their derivations.
%It is noted that we describe longer equations in the supplementary notebook, which is separately prepared from the main manuscript.

\section{Bases of higher-order curvature gravity with a scalar field}\label{S2}

The action of $f(R,G)$ gravity is examined in this section, where $f(R,G)$ is an arbitrary function of the Gauss-Bonnet and the Ricci scalar fields.  Emphasizing that $f(R,G)$ theory is an extension of GR and that it coincides with GR when $f(R,G)=R+G$, and we have an alternative theory to GR when $f(R)\neq R+G$.  In $D$-dimension, we begin with the action of $f(R,G)$ gravity as indicated by\footnote{ In this work, we focus on the situation $f(R,G)=f(R)+f_1(G)$ to derive a model that is ghost-free by expressing $f_1(G)$ as $H(\phi_1)G$, where $\phi_1$ is a scalar field (for more information, see Refs.~\cite{Nojiri:2024hau,Nojiri:2023qgd,Nashed:2022mij})}
\begin{align}
\label{g2}
{S}=\int d^D x \sqrt{-g}\left\{ \frac{1}{2\kappa^2}f(R)
 - \frac{1}{2} \partial_\mu \phi_1 \partial^\mu \phi_1+V(\phi_1)+ H(\phi_1) {G} \right\}\, .
\end{align}
Here, $\phi_1$ is the scalar field and $V(\phi_1)$ is the potential for $\phi_1$ and $H(\phi_1)$ is also a function of $\phi_1$. Furthermore, the Gauss-Bonnet invariant $G$ is defined by
\begin{align}
\label{eq:GB}
{G} = R^2-4R_{\alpha \beta}R^{\alpha \beta}+R_{\alpha \beta \rho \sigma}R^{\alpha \beta \rho \sigma}\,.
\end{align}
Despite being total-derivative in four dimensions ($D=4$), the term containing the Gauss-Bonnet invariant contributes non-trivial amounts to the system's field equations because of the coupling $H(\phi_1)$.
 In order to prevent the fifth force from showing up, we additionally suppose that the matters do not couple with the scalar field $\phi_1$.

When the action (\ref{g2}) is varied w.r.t. the scalar field $\phi_1$, it gives
%the following equation
\begin{align}
\label{g3}
\nabla^2 \phi_1-V'(\phi_1)+ H'(\phi_1){G}=0\, .
\end{align}
Additionally, by varying the action (\ref{g2}) w.r.t.  the metric $g_{\mu\nu}$, we get
%the following equations
\begin{align}
\label{GBeq}
&0= -\frac{1}{2\kappa^2}\left(R_{\mu \nu} \mathit{f_{R}}-\frac{g_{\mu \nu}}{2}f(R)+[g_{\mu \nu}\Box -\nabla_\mu \nabla_\nu]f_R\right)+\frac{1}{2} \partial^\mu \phi_1 \partial^\nu \phi_1
 - \frac{1}{4}g^{\mu\nu} \partial_\rho \phi_1 \partial^\rho \phi_1+ \frac{ g^{\mu\nu}}{2}[H(\phi_1)G -V(\phi_1)]
+2 H(\phi_1) R R^{\mu\nu} \nonumber \\
& + 2 \nabla^\mu \nabla^\nu \left(H(\phi_1)R\right)- 2 g^{\mu\nu}\nabla^2\left(H(\phi_1)R\right)
+ 8H(\phi_1)R^\mu_{\rho} R^{\nu\rho}- 4 \nabla_\rho \nabla^\mu \left(H(\phi_1)R^{\nu\rho}\right)
 - 4 \nabla_\rho \nabla^\nu \left(H(\phi_1)R^{\mu\rho}\right)+ 4 \nabla^2 \left( H(\phi_1) R^{\mu\nu} \right) \nonumber \\
& + 4g^{\mu\nu} \nabla_{\rho} \nabla_\sigma \left(H(\phi_1) R^{\rho\sigma} \right)
 - 2 H(\phi_1) R^{\mu\rho\sigma\tau}R^\nu_{ \rho\sigma\tau}+ 4 \nabla_\rho \nabla_\sigma \left(H(\phi_1) R^{\mu\rho\sigma\nu}\right)\, .
\end{align}
By using the following Bianchi identities, we obtain
\begin{align}
\label{Bianchi}
&\nabla^\rho R_{\rho\tau\mu\nu}= \nabla_\mu R_{\nu\tau} - \nabla_\nu R_{\mu\tau} \, , \qquad
\nabla^\rho R_{\rho\mu} = \frac{1}{2} \nabla_\mu R\, , \qquad
\nabla_\rho \nabla_\sigma R^{\mu\rho\nu\sigma} =\nabla^2 R^{\mu\nu} - \frac{1}{2}\nabla^\mu \nabla^\nu R
+ R^{\mu\rho\nu\sigma} R_{\rho\sigma}- R^\mu_{\ \rho} R^{\nu\rho}\, , \nonumber \\
&\nabla_\rho \nabla^\mu R^{\rho\nu} + \nabla_\rho \nabla^\nu R^{\rho\mu}
= \frac{1}{2} \left(\nabla^\mu \nabla^\nu R + \nabla^\nu \nabla^\mu R\right)
 - 2 R^{\mu\rho\nu\sigma} R_{\rho\sigma} + 2 R^\mu_{\ \rho} R^{\nu\rho}\, , \qquad
\nabla_\rho \nabla_\sigma R^{\rho\sigma} = \frac{1}{2} \Box R \, ,
\end{align}
in Eq.~(\ref{GBeq}), we find
\begin{align}
\label{gb4b}
0=&\, -\frac{1}{2\kappa^2}\left(R_{\mu \nu} {f_{R}}-\frac{1}{2}g_{\mu \nu}f(R)+[g_{\mu \nu}\Box -\nabla_\mu \nabla_\nu]f_R\right)
+ \left(\frac{1}{2} \partial^\mu \phi_1 \partial^\nu \phi_1
 - \frac{1}{4}g^{\mu\nu} \partial_\rho \phi_1 \partial^\rho \phi_1 \right)
+ \frac{1}{2} g^{\mu\nu} \left[ H(\phi_1) G-V(\phi_1) \right] \nonumber \\
&\, -2 H(\phi_1) R R^{\mu\nu} + 4H(\phi_1)R^\mu_{\ \rho} R^{\nu\rho}
 -2 H(\phi_1) R^{\mu\rho\sigma\tau}R^\nu_{\ \rho\sigma\tau}
 -4 H(\phi_1) R^{\mu\rho\sigma\nu}R_{\rho\sigma}+ 2 \left( \nabla^\mu \nabla^\nu H(\phi_1)\right)R- 2 g^{\mu\nu} \left( \nabla^2H(\phi_1)\right)R\nonumber \\
&\,  - 4 \left( \nabla_\rho \nabla^\mu H(\phi_1)\right)R^{\nu\rho}
 - 4 \left( \nabla_\rho \nabla^\nu H(\phi_1)\right)R^{\mu\rho} + 4 \left( \nabla^2 H(\phi_1) \right)R^{\mu\nu}
+ 4g^{\mu\nu} \left( \nabla_{\rho} \nabla_\sigma H(\phi_1) \right) R^{\rho\sigma}
 - 4 \left(\nabla_\rho \nabla_\sigma H(\phi_1) \right) R^{\mu\rho\nu\sigma}.
\end{align}
In the 4-dimension case, namely, when $D=4$, Eq.~(\ref{gb4b}) is reduced to
\begin{align}
\label{gb4bD4}
{\mathop{{ I}}}_{\mu \nu}= & -\frac{1}{2\kappa^2}\left(R_{\mu \nu} \mathit{f_{R}}-\frac{1}{2}g_{\mu \nu}f(R)+[g_{\mu \nu}\Box -\nabla_\mu \nabla_\nu]f_R\right)
+ \left(\frac{1}{2} \partial^\mu \phi_1 \partial^\nu \phi_1
 - \frac{1}{4}g^{\mu\nu} \partial_\rho \phi_1 \partial^\rho \phi_1 \right)
 - \frac{1}{2} g^{\mu\nu}V(\phi_1) \nonumber \\
& + 2 \left( \nabla^\mu \nabla^\nu H(\phi_1)\right)R
 - 2 g^{\mu\nu} \left( \nabla^2H(\phi_1)\right)R
 - 4 \left( \nabla_\rho \nabla^\mu H(\phi_1)\right)R^{\nu\rho}
 - 4 \left( \nabla_\rho \nabla^\nu H(\phi_1)\right)R^{\mu\rho}+ 4 \left( \nabla^2 H(\phi_1) \right)R^{\mu\nu} \nonumber \\
& + 4g^{\mu\nu} \left( \nabla_{\rho} \nabla_\sigma H(\phi_1) \right) R^{\rho\sigma}
- 4 \left(\nabla_\rho \nabla_\sigma H(\phi_1) \right) R^{\mu\rho\nu\sigma}=0.
\end{align}

The field equations (\ref{gb4bD4}) yields the trace in the following form
\begin{eqnarray} \label{f3}
{\mathop{{ I}}}=3\Box f_R+\mathit{\mathit{ R}}f_R-2f(R)+\frac{1}{2}\partial^\rho \phi_1 \partial_\rho \phi_1+2V(\phi_1)+2\left( \nabla^2 H(\phi_1)\right)R-4 \left( \nabla_\rho \nabla_\sigma H(\phi_1) \right)R^{\rho\sigma} \equiv0 \,.\end{eqnarray}
The following form of $ f( R )$ may be obtained from Eq.~(\ref{f3}) as\footnote{Due to the nature of the present study, where we are interested to find a vacuum solution, we use the relativistic units that enable us to put $2\kappa^2=1$.}
\begin{eqnarray} \label{f3s}
f(R)=\frac{1}{2}\Big[3\Box f_R+Rf_R+\frac{1}{2}\partial^\rho \phi_1 \partial_\rho \phi_1+2V(\phi_1)+2\left( \nabla^2 H(\phi_1)\right)R-4 \left( \nabla_\rho \nabla_\sigma H(\phi_1) \right)R^{\rho\sigma}\Big]\,.\end{eqnarray}
By combining Eq.~(\ref{f3s}) and Eq.~(\ref{gb4bD4}), we have
\begin{eqnarray} \label{f3ss}
{\mathop{{ I}}}_{\mu \nu}&=&R_{\mu \nu} f_R-\frac{1}{4}g_{\mu \nu} { R}f_R+\frac{1}{4}g_{\mu \nu}\Box f_R -\nabla_\mu \nabla_\nu f_R+\frac{1}{8}g_{\mu \nu}\partial^\rho \phi_1 \partial_\rho \phi_1-\frac{1}2\partial_\mu \phi_1 \partial_\nu \phi_1-2 \left( \nabla^\mu \nabla^\nu H(\phi_1)\right)R
 \nonumber\\&+&\frac{3}2 g^{\mu\nu} \left( \nabla^2H(\phi_1)\right)
  + 4 \left( \nabla_\rho \nabla^\mu H(\phi_1)\right)R^{\nu\rho}
 +4 \left( \nabla_\rho \nabla^\nu H(\phi_1)\right)R^{\mu\rho}- 4 \left( \nabla^2 H(\phi_1) \right)R^{\mu\nu} \nonumber\\&- &3g^{\mu\nu} \left( \nabla_{\rho} \nabla_\sigma H(\phi_1) \right) R^{\rho\sigma}
+4 \left(\nabla_\rho \nabla_\sigma H(\phi_1) \right) R^{\mu\rho\nu\sigma} =0   \,.
\end{eqnarray}
The Gauss-Bonnet invariant is known to be a four-dimensional total derivative.  Consequently, the component containing the Gauss-Bonnet invariant does not contribute to Eq.~(\ref{f3ss}) if $H$ is a constant.
 The following section will cover the application of the field equation \eqref{f3ss} to a spacetime that is spherically symmetric and attempt to derive an analytic solution to the output differential equations.

%%%%%%%%%%%%%%%%%%%%%%%%%%%%%%%%%%% Section 3 %%%%%%%%%%%%%%%%%%%%%%%%%%%%%%%%%%%%%%%%
\section{Spherically symmetric black hole solution}\label{S3}
%%%%%%%%%%%%%%%%%%%%%%%%%%%%%%%%%%%%%%%%%%%%%%%%%%%%%%%%%%%%%%%%%%%%%%%%%%%%%%%%%%%%%%
   In order to study the equations of motion presented in Eq.~(\ref{f3ss}), and provide the forms of arbitrary functions, $f(R)$ and $H(\phi_1)$,  without introducing any additional constraints on  $R$ or  $H(\phi_1)$, we utilize a spherically symmetric spacetime framework. This spacetime is characterized by two unknown functions, which are expressed in a specific form to facilitate our analysis. This approach allows us to explore the full range of possibilities for $f(R)$ and $H(\phi_1)$, thereby providing a comprehensive view of the solutions within the context of this modified gravity theory. The line-element considered in this study is expressed as follows
%%%%%%%%%%%%%%%%%%%%%%%%%%%%%%%%%%% Section 3 %%%%%%%%%%%%%%%%%%%%%%%%%%%%%%%%%%%%%%%%
%\subsection{Spherically symmetric solution}
%Assuming the spherically-symmetric the line-element to be in the form:
\begin{eqnarray} \label{met12}
& &  ds^2=-S(r)dt^2+\frac{dr^2}{S_1(r)}+r^2(d\theta^2+\sin^2\theta d\phi^2)\,,  \end{eqnarray}
where the functions $S(r)$ and $S_1(r)$ rely only on $r$.  From the metric (\ref{met12}), the Ricci scalar reads
  \begin{eqnarray} \label{Ricci}
    R=\frac{r^2S_1S'^2-r^2SS'S'_1-2r^2SS_1S''-4rS[S_1S'-SS_1']+4S^2(1-S_1)}{2r^2S^2}\,,
  \end{eqnarray}
  where $S\equiv S(r)$,  $S\equiv S(r)$,  $S'=\frac{dS}{dr}$, $S''=\frac{d^2S}{dr^2}$ and $S'=\frac{dS}{dr}$.
By plugging Eqs.~(\ref{f3s}), (\ref{f3ss}) with Eq.~(\ref{met12}) and using Eq.~(\ref{Ricci}),  we acquire the components of the gravitational field equation \eqref{f3ss}, whose expressions are described in Eq.~\eqref{feq} of Appendix~{\color{blue}A}. %\footnote{Here, we mention that all explicit forms of the field equation components are included in the Appindex.}.
Since spherical symmetry is taken into account in this study, we can assume that $f(R)$ depends only on   $r$, i.e., $f(R)=f(r)$.  It is worth noting that equations in \eqref{feq} are identical with those presented in Refs.~\cite{Jaime:2010kn,Nojiri:2023qgd} when  $H(r)=1$. If $F(r)=1$ and $H(r)=constant$, then we can show that Eqs.~\eqref{feq}, except for the trace one, are identical to those presented in Refs.~\cite{Nashed:2021lzq,Nashed:2020mnp}). If $H(r)\neq constant$ and $F(r)\neq 1$, then we have three differential equations, except the trace one, in five unknown ones, $S,\, S_1,\, F,\, H,\,
\phi_1$. To close such a system, we need two extra conditions.

The trace equation in~\eqref{feq} shows that for $F=H(r)=V(r)=0$, we get $\textit{f(R)=R}$ and $S=S_1$ leading to the Schwarzschild geometry. Hence, the functions $F$ and $H(r)$ play a pivotal role in this study only when they have a non-vanishing value. For the case that $S(r)$ and $S_1(r)$ are arbitrary, the forms of $F$, $H(r)$ and $\phi(r)$ can be derived.
The operation that the first equation minus the third equation in~\eqref{feq} of Appendix~{\color{blue}A} yields
\begin{align}\label{F}
&F(r)=\frac{1}{\Upsilon}\left[ \int \frac{ \left\{ 4S_1   \left\{ S_1   S' r-2S   \left(S_1-1   \right)  \right\} S  H'' + \left[  4S_1{}^{2}S  S''r-2S_1{}^{2} S'^{2}r+6S_1  S'r S'_1S  -4S^{2} \left( 3S_1  -1\right) S'_1   \right]H'   \right\}\Upsilon}{ S_1 S r \left( rS' -2S    \right)}dr+c_1 \right]\,,
\end{align}
where $\Upsilon\equiv \Upsilon(r)$ is defined as
\begin{align}\label{FF2}
\Upsilon(r)= \exp\left[ \int \frac {2\,S_1   S'' S  {r}^{2}-S_1   S'^{2}{r}^{2}+rS   \left(  S'_1  r+2S_1   \right) S'  -2 S^{2} \left(2S_1  + S'_1 r -2\right) }{2S_1  S  r \left( rS'-2S   \right) }{dr}\right]\,.
\end{align}
%It follows from the fact that
With $F(r)\equiv f_R=\frac{\partial f}{\partial r}\frac{\partial r}{\partial R}$, the form of $\textit{f(R)}\equiv f(r)$ is represented as
\begin{align}\label{f3}
&f(r)=\int -\frac{1}{2\Upsilon_1 S^{3}{r}^{3}} \left[ \int\frac{2\Upsilon_1 }{ S_1 S{r}\left( 2S  -rS'  \right)} \left\{ 6 H'S'_1  S^{2}S_1  -4 H'' S_1   S^{2}+4 H''  S_1{}^{2} S^{2}-2 H''  S_1{}^{2}S  rS'  -2  H'  S'_1  S^{2}-2 H' S_1{}^{2}S  S'' r\right.\right.\nonumber\\
&\left.\left.+ H'  S_1{}^2S'^{2}r-3 H' S_1  S'r S'_1 S \right\}{dr}+c_1 \right] \left[3 S''S'_1  S^{2}{r}^{3} -2 S'^{2}  S'_1  S  {r}^{3}+  S'   S''_1 S^{2}{r}^{3}-4S_1   S''  S' S  {r}^{3}+2S_1   S''' S ^{2}{r}^{3}+2S_1   S'^{3}{r}^{3}\right.\nonumber\\
&\left.-4S_1   S'^{2}S  {r}^{2}-4 S^{2}r  S' S_1  +4S'_1  S'  S^{2}{r}^{2}+4S_1  S''  S^{2}{r}^{2}+4 S''_1  S^{3}{r}^{2}+8 S^{3}-8 S^{3}S_1  \right] {dr}+c_2\,,
\end{align}
where $\Upsilon_1\equiv \Upsilon_1(r)$ is defined as
\begin{align}\label{f33}
\Upsilon_1=\exp\left[ \int \!{\frac {-2 \,S_1    S'S  r+S_1   S'^{2}{r}^{2}-2\,S_1   S'' S  {r}^{2}+4\,S_1  S^{2}+2\,S'_1  S^{2}r-4\, S^{2}- S' S'_1  S {r}^{2}}{2S_1  S  r \left( 2\,S  -rS'   \right) }}{dr} \right]\,.
\end{align}
The calculation that the first equation plus 1/3 times the second equation in~\eqref{feq} leads to
\begin{align}\label{H}
&H(r)=\int  \left[ \int\frac{\Upsilon_2}{S_1 } \left\{ F  S  {r}^{2}S_1 S''  -\frac{F  {r}^{2}S_1   S'^{2}}2+\frac{1}2 S  r \left\{ rF  S'_1 +2S_1   \left( F  +rF'   \right)  \right\}S'  - S^{2} \left\{ rF  S'_1  + 2 F'S_1  r+2F   \left( S_1-1 \right)  \right\}  \right\}\right.\nonumber\\
  &\left.\left\{6 S_1  S''S  {r}^{2}-3S_1   S'^{2}{r}^{2}+S  r \left(16S_1  +3 S'_1 r \right) S'  +6 S^{2} \left( {r}^{2}+4 S_1  -4+12 S'_1 r \right)  \right\} ^{-1}{dr}+4c_1\right] \Upsilon_2{}^{-1}{dr}+c_2\,,
\end{align}
where $\Upsilon_2\equiv \Upsilon_2(r)$ is defined as
\begin{align}
&\Upsilon_2=\exp\left\{\frac{1}2\,\int\left\{ \left[ 6S_1{}^{2}S  {r}^{3}S'  +2{r}^{2}S_1   \left\{ 16S_1  + S'_1 r \right\}  S^{2} \right] S''  -3S_1^{2} S'^{3}{r}^{3}-4\, S_1{}^{2} S'^{2}{r }^{2}S  +3 S^{2}r S'  \left[ S'_1{}^{2}{r}^{2}+16S_1  S'_1r\right.\right.\right.\nonumber\\
&\left.\left.\left.+2S_1   \left( {r}^{2}+10\,S_1 -2  \right)  \right]  +2S^{3} \left[ 6S'_1{}^{2}{r}^{2}+r \left( 18S_1 +3{r}^{2}-2 \right) S'_1  +12S_1   \left( 2S_1  +{r}^{2} -2\right)  \right] \right\}\left\{ \left[ 6S_1  S''S  {r}^{2 }-3S_1 S'^{2}{r}^{2}\right.\right.\right.\nonumber\\
&\left.\left.\left. +S  r\left( 16S_1  +6 S'_1r \right)S'  +2S^{2} \left\{3 {r}^{2}+2S_1 -2+2 S'_1 r \right\}  \right] S  rS_1  \right\}^{-1}{dr}\right\}\,.
\end{align}
Finally, the procedure that the second equation minus the first one in~\eqref{feq} gives
\begin{align}\label{p}
&\phi_1=\pm\int \frac {1}{S_1  S  r}\left\{-S_1  S   \left[ 12H'  S_1{}^{2}S'  -12H' S'_1 S  S_1  -4 H' S_1  S'  +8H''  S_1  S  +2S  {r}^{2} F'' S_1  +S  {r}^{2} F'  S'_1  +2F S  rS'_1  -2F  rS_1 S' \right.\right.\nonumber\\
 &\left.\left.-8 H'' S_1{}^{2}S  +4H' S'_1 S  -{r}^{2} F'   S_1  S'  \right] \right\}^{1/2}{dr}+c_1\,.
\end{align}
It follows from Eq.~$\eqref{p}$ that the expression under square root must be positive, namely,
\begin{eqnarray}\nonumber
&& -S_1  S  \bigg[ 12H'  S_{1}^{2}S'  -12H' S'_1 S  S_1  -4 H' S_1  S'  +8H''  S_1  S  +2S  {r}^{2} F'' S_1  +S  {r}^{2} F'  S'_1  +2F S  rS'_1  -2F  rS_1 S'\\\nonumber&&
-8 H'' S_1{}^{2}S  +4H' S'_1 S  -{r}^{2} F'   S_1  S' \bigg]>0,
\end{eqnarray}
otherwise, ghosts will appear.
In Eqs.~\eqref{F}, \eqref{H} and \eqref{p}, $c_1$ and $c_2$ are constants of integration. In the following subsection, we take certain forms of the ansatzs $S$ and $S_1$ that modifies the Schwarzschild solution.
  %Using Eq. \eqref{sol} in Eq. \eqref{feq} we get a lengthy form of $H(r)$, $\phi_1$ and $V$ which we  list them in Appendix A
%\begin{align}
%\end{align}

%%%%%%%%%%%%%%%%%%%%%%%%%%%%%%%%%%% Section 3 %%%%%%%%%%%%%%%%%%%%%%%%%%%%%%%%%%%%%%%%
\subsection{Modified Schwarzschild solution }
%%%%%%%%%%%%%%%%%%%%%%%%%%%%%%%%%%%%%%%%%%%%%%%%%%%%%%%%%%%%%%%%%%%%%%%%%%%%%%%%%%%%%%
In this study, we take the ansatz for $S$ and $S_1$ that modify the Schwarzschild spacetime as\footnote{ The form of the ansatzs of Eq.~\eqref{sol} deform the Schwarzschild spacetime. As the dimensional constant  $\digamma$ is equal to zero, we recover the Schwarzschild spacetime. So, we are going to use the ansatzs of Eq.~\eqref{sol} as an input in the equations Eqs.~\eqref{F}, \eqref{H}, and \eqref{p} and study the effect of the modified theory $f(R,G)$.}
\begin{align}\label{sol}
S=1-\frac{c}{r}+\frac{\digamma}{r^2}\,,\qquad
S_1=1-\frac{c}{r}+\frac{c\digamma^{3/2}}{r^4}\,,
\end{align}
where $c=2M$ is the gravitational mass of the system and $\digamma$ is a dimensional parameter that has a unit of ${\textit lenth^2}$ and is responsible to modify the Schwarzschild spacetime. It is seen that if $\digamma=0$, the Schwarzschild geometry can be recovered.
 By plugging Eq.~\eqref{sol} into Eqs.~\eqref{F},~\eqref{H}, and \eqref{p}, we obtain lengthy expressions for $H(r)$, $F(r)$, $f(r)$ etc. In Appendix {\color{blue} A}, we list how one can derive these functions, $H(r)$, $F(r)$, $f(r)$ etc.  Here, we present the asymptotic form of the functions, $\phi_1(r)$, $F(r)$, $f(r)$, $V(r)$ which yield the following behavior as $r\to \infty$
\begin{align}
&F(r)_{_{_{r\to \infty}}}\approx {\frac {c_2\,c \left( 70\,\digamma{c}^{3}-240\,{
\digamma}{c}^{2}-64\,{\digamma}^{3/2}c+96\,{\digamma}^{2} \right) }{64{\digamma}^{5/2}}}+{\frac {c_2\,c \left( 80\,{c}^{3}-192\,{\digamma}c-128\,{\digamma}^{3/2} \right) }{64{\digamma}^{3/2}r}}+{\frac {c_2\,c \left( 96{c}^{2}-128\,{\digamma} \right) }{64{\digamma}^{1/2
}{r}^{2}}}\,,\nonumber\\
&H(r)_{_{_{r\to 0}}} \approx c_1+\frac{c_2\,{r}^{2}}2+\frac {c_2\,c{r}^{3}}{6
\digamma}-\frac { \left( 4\,\digamma -3\,{c}^{2}\right) c_2\,{r}^{4
}}{32\digamma^{2}}+{\frac {1}{80}}\,{\frac { \left( 5\,{c}^
{3}\digamma^{15/2}-12\,\digamma^{17/2}c+8\,\digamma^{9} \right) c_2\,{r}^{5}}{\digamma^{21/2}}},\nonumber\\
&f(r)_{_{_{r\to \infty}}} \approx -\frac {{c}^{2}c_2\, \left( 15435\,{c}^{4}-15120\,\digamma^{3/2}
c-24360\,{c}^{2}\digamma-1680\,\digamma^2
 \right) }{5040\digamma{r}^{6}}-\frac{{c}^{2}c_2\, \left( 8640\,\digamma^{5/2}-17280\,\digamma^{2}c+17280\,\digamma{c}^{3}
 \right) }{5040\digamma{r}^{7}},\nonumber\\
%&-{\frac {1}{5040}}\,{\frac {{c}^{2}c_2\, \left( -5040\,\digamma^{7/2}+16380\,\digamma^{5/2}{c}^{2} \right) }{\digamma^{3/2}{r}^{8}}}-{\frac {16}{9}}\,{\frac {\digamma^{2}{c}^{3}c_2}{{r}^{9}}}+{\frac {24}{5}}\,{\frac {\digamma^{3}{c}^{2}c_2}{{r}^{10}}}\,,\nonumber\\
&\phi_1(r)_{_{_{r\to \infty}}} \approx
{\frac {-256\sqrt {m}\ln  \left( r
 \right) \sqrt {-c_2}\sqrt {3}\digamma+187\sqrt {-c_2}\ln  \left( r \right) {m}^{5/2}\sqrt {3}+768c_1
\,\digamma^{5/4}}{768\digamma^{5/4}}}+{\frac {3{m}^{3/2}\sqrt {-c_2}\sqrt {3
}}{4\sqrt [4]\digamma r}}+{\frac {4\sqrt {3c}\digamma^{3/4}\sqrt {-c_2}}{{r}^{2}}}\,,\nonumber\\
&V(r)_{_{_{r\to \infty}}}\approx C_1+\frac{C_2}{r}+\frac{C_3}{r^2}\,,
\label{scal1}\end{align}
where $C_1$, $C_2$ and $C_3$ are constants depend on $c$, $c_1$, $c_2$ and $\digamma$.
The pattern of the scalar field, $\phi_1(r)$, as give by Eq.~\eqref{scal1} is illustrate in  Fig.~\ref{Figg:1} \subref{fig:1a}
shows that it remains positive throughout. In the following section, we will analyze the thermodynamical quantities and physical characteristics of the black hole, as determined by Eq.~\eqref{sol}.

%%%%% Sec. IV %%%%%
\section{Physics of the black hole
%(\ref{sol})
}\label{iv}
In this section, we analyze the various phases of the solution given by Eq.~\eqref{sol}, focusing on the asymptotic behavior of the scalar invariants and their associated thermodynamic properties.
\subsection{Singularity test}
When $c=2M$, the line element~(\ref{met12}) after using Eq. (\ref{sol})  expressed as,
\begin{eqnarray}\label{met3} ds^2=-\left(1-\frac{2M}{r}+\frac{\digamma }{r^2}\right)dt^2+\left(1-\frac{2M}{r}+\frac{2M\digamma^{3/2}}{r^4}\right)^{-1} \ dr^2+r^2(d\theta^2+\sin^2d\phi^2). \end{eqnarray}
It is simple to verify that a smooth transition back to the Schwarzschild spacetime occurs when the value of $\digamma$ is equal to zero \cite{Misner:1974qy}.

 To understand the physics of the BH given by Eq.~(\ref{met3}) within $f(R,G)$ theory, we are going to describe its relevant scalars, including the Kretschmann invariant, the Ricci tensor, etc., in the limit of $r \to \infty$. By carrying out such calculations, we get
\begin{eqnarray}\label{ST1}
&&R(r)_{r \to \infty}\approx {\frac {-{m}^{2}-4\digamma}{2{r}^{4}}}+{\frac {2m\digamma-{m}^{3}}{{r}^{5}}},
+{\frac {8\digamma{m}^{2}-3{m}^{4}+12m{\digamma}^{3/2}+8\,{\digamma}^{2}}{2{r}^{6}}},\nonumber\\
%-2{\frac {\digamma}{{r}^{4}}}-{\frac {\digamma c}{{r}^{5}}}+{\frac {6{\digamma}^{3/2}c+7
% \left( -3/14{c}^{2}+4/7\digamma \right) \digamma}{{r}^{6}}}\,,\nonumber\\
&& \left(R_{\mu \nu} R^{\mu \nu}\right)_{ r\to \infty} \approx 12{\frac {{\digamma}^{2}}{{r}^{8}}}+12r{\frac {{\digamma}^{2}c}{{r}^{9}}}+
 \frac{\left( -20r{\digamma}^{5/2}c-32r{\digamma}^{2} \left( \digamma-{\frac {17}{32}}r{c}^{2}
 \right)  \right)} {r^{10}}\,,\nonumber\\
  && \left(R_{\mu \nu \rho \sigma} R^{\mu \nu \rho \sigma}\right)_{ r\to \infty} \approx{\frac {12{c}^{2}}{{r}^{6}}}-{\frac {32\digamma c}{{r}^{7}}}+{\frac {44\,{\digamma}^{2}-16\digamma{c}^{2}}{{r}^{8}}}+{\frac {84{\digamma}^{2}c-18\digamma{c}^{3}-24
{c}^{2}{\digamma}^{3/2}}{{r}^{9}}}\,,\nonumber\\
&&G(r)_{r \to \infty}\approx 12{\frac {{c}^{2}}{{r}^{6}}}-32{\frac {\digamma c}{{r}^{7}}}-16{\frac {\digamma
{c}^{2}}{{r}^{8}}}+2{\frac {c \left( -9\digamma{c}^{2}+20{\digamma}^{2}-12{\digamma
}^{3/2}c \right) }{{r}^{9}}}\,.\nonumber\\
\end{eqnarray}
Here, $R$ is the Ricci scalar, $R^{\alpha\beta}R_{\alpha\beta}$ is the Ricci squared invariant, $R^{\mu\nu\alpha\beta}R_{\mu\nu\alpha\beta}$  is the Kretschmann scalar, and  $G$ denotes the Gauss-Bonnet term.

\begin{figure*}
\centering
\subfigure[~The behavior of the scalar field $\phi_1$ as a function of the radial coordinate $r$]{\label{fig:1a}\includegraphics[scale=0.25]{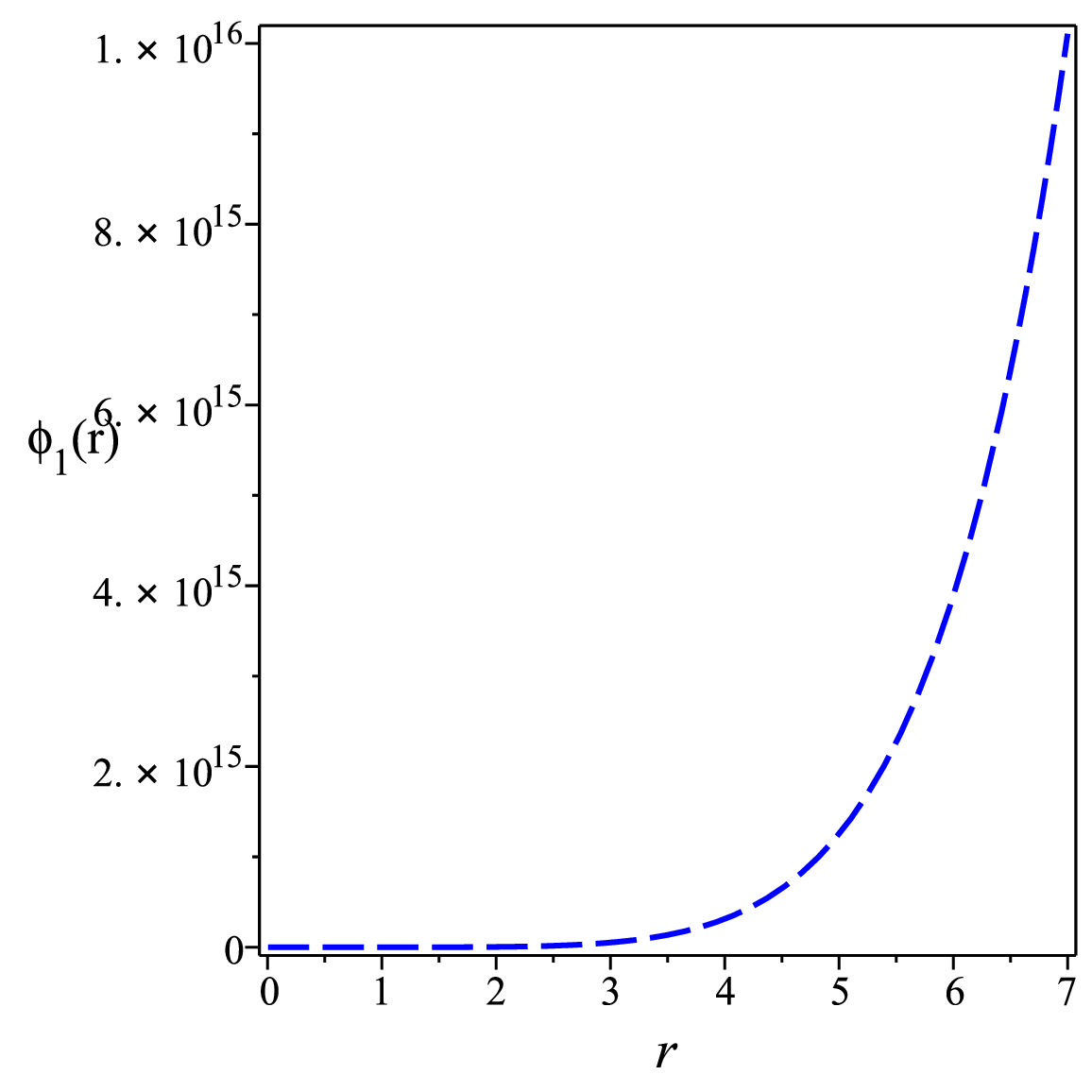}}\hspace{0.5cm}
\subfigure[~The variation of the Ricci scalar $R$ with respect to the radial coordinate $r$]{\label{RS1}\includegraphics[scale=0.25]{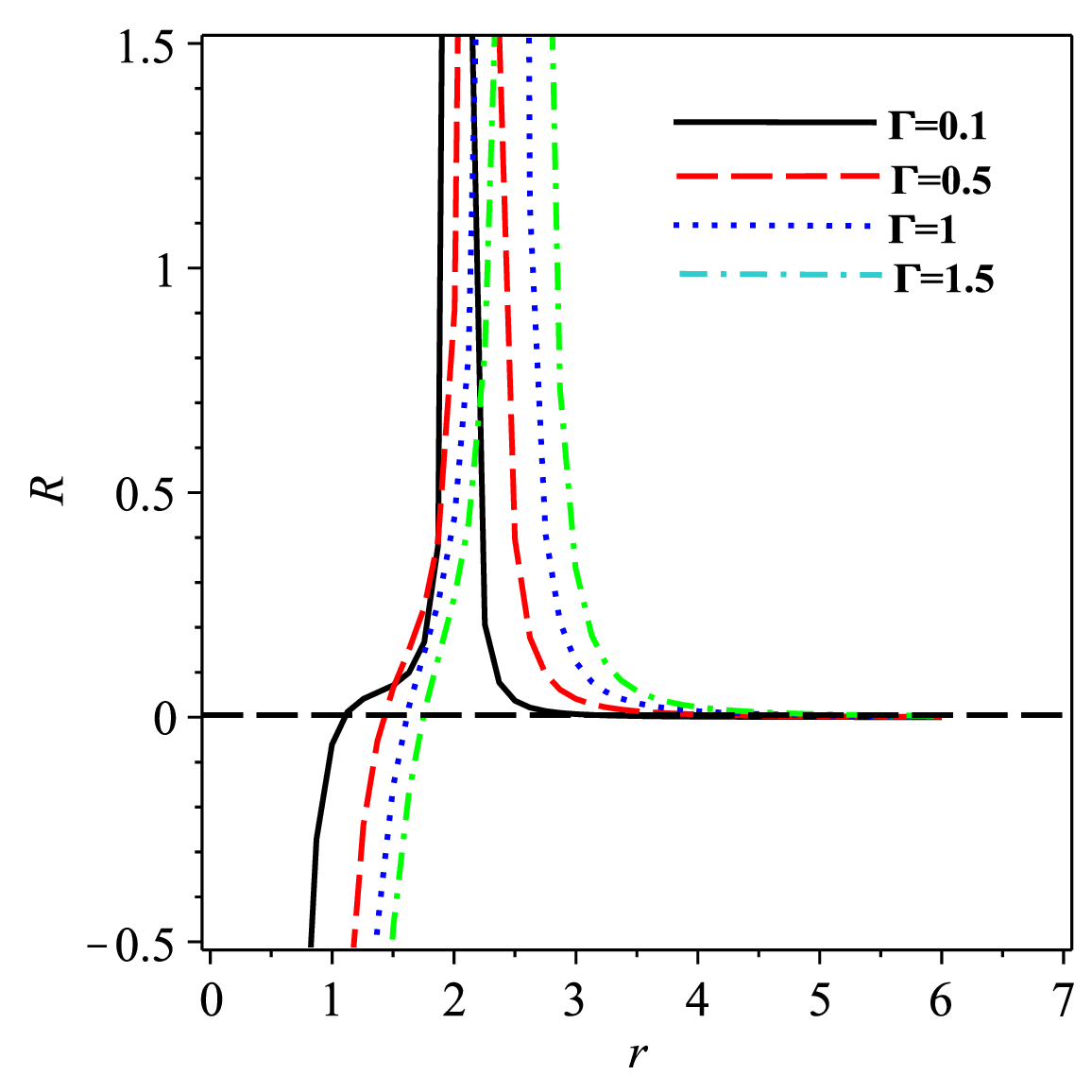}}\hspace{0.45cm}
\subfigure[~The variation of the Ricci square $R^{\alpha\beta}R_{\alpha\beta}$ with respect to the radial coordinate $r$.]{\label{RS2}\includegraphics[scale=0.25]{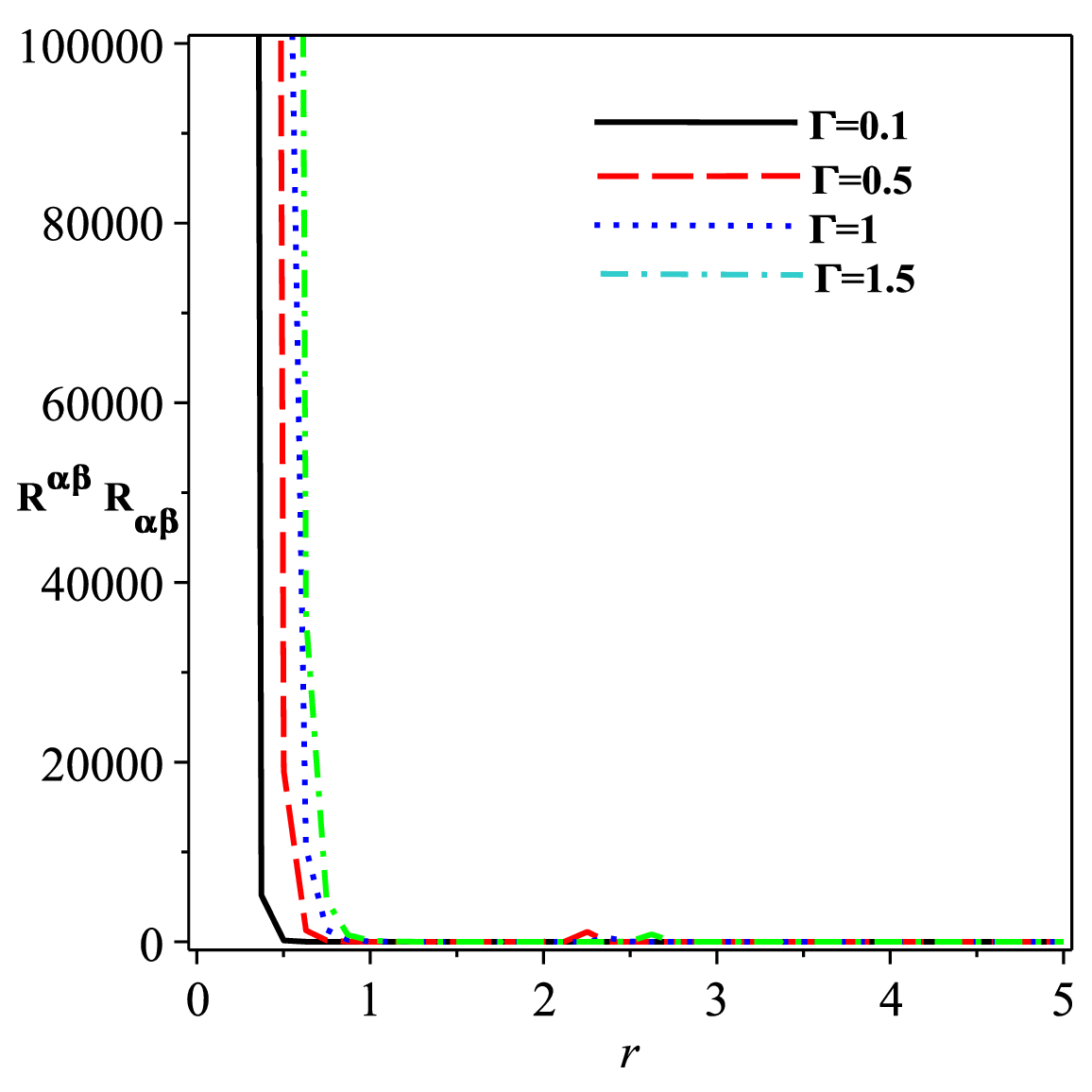}}\hspace{0.45cm}
\subfigure[~The variation of the Kretschmann scalar $K=R^{\mu\nu\alpha\beta}R_{\mu\nu\alpha\beta}$ with respect to the radial coordinate $r$.]{\label{RS3}\includegraphics[scale=0.25]{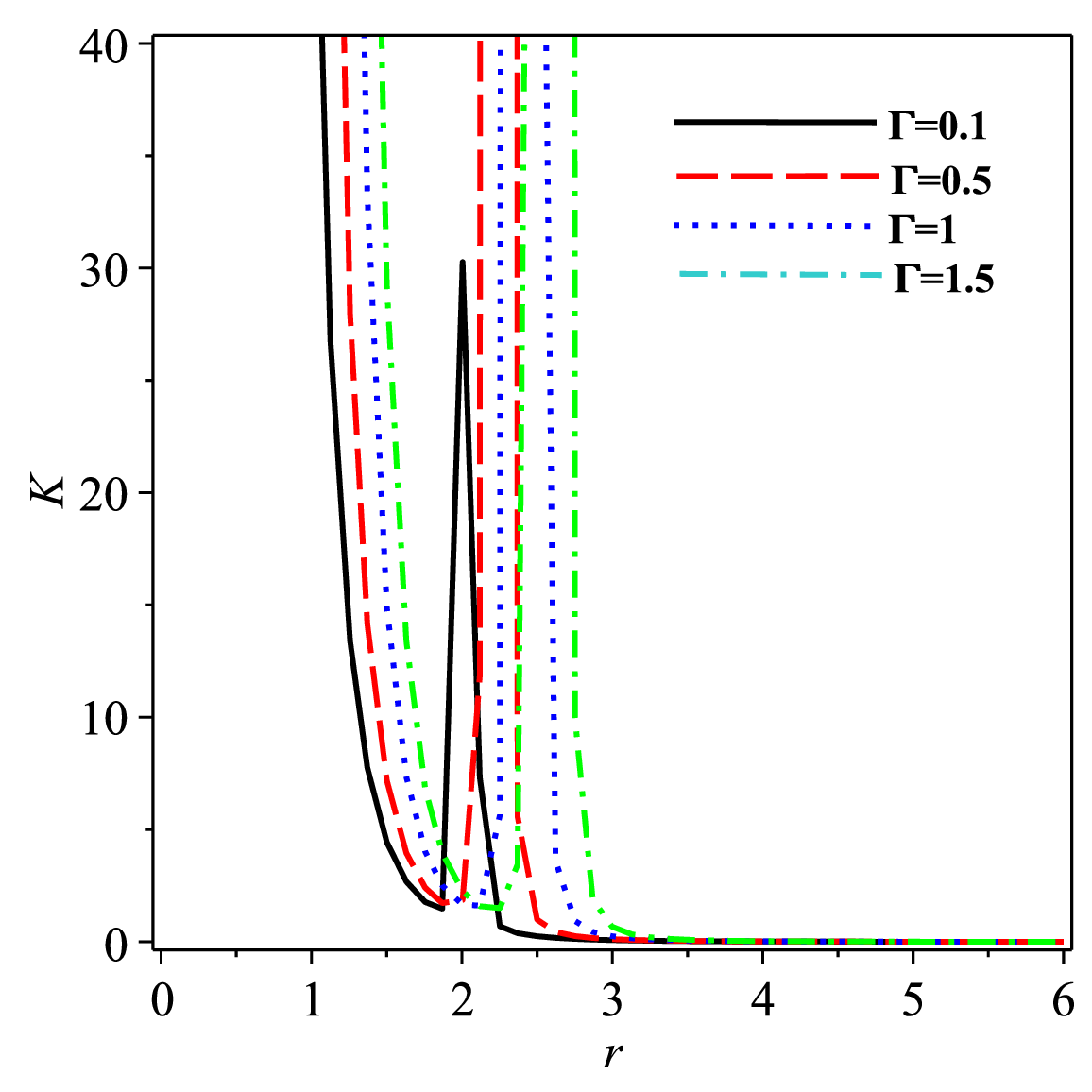}}\hspace{0.45cm}
\subfigure[~The behavior of the potential $V$ as a function of the scalar field]{\label{fig:1b}\includegraphics[scale=0.25]{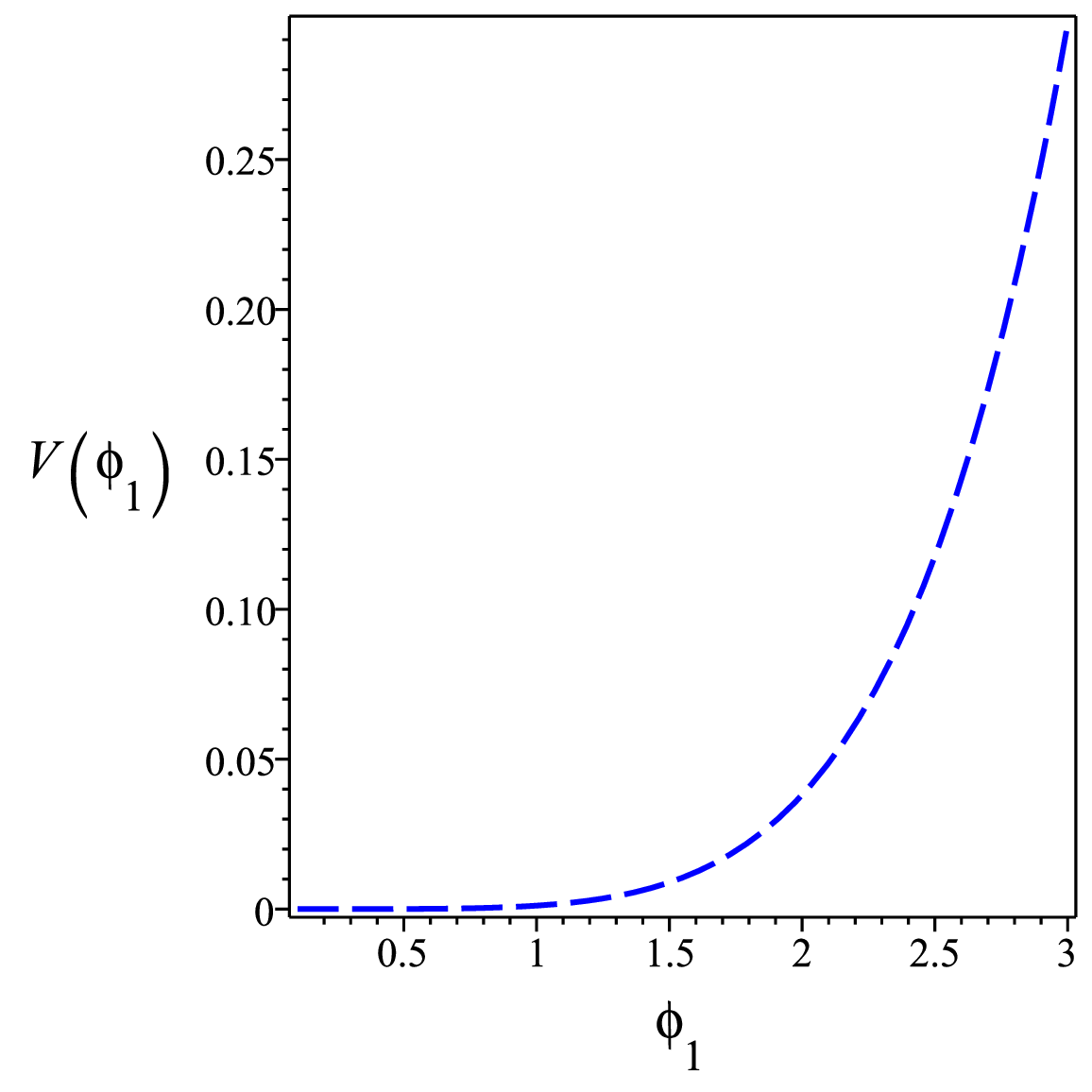}}\hspace{0.5cm}
\subfigure[~The behavior of the function $f(R)$]{\label{fig:1c}\includegraphics[scale=0.25]{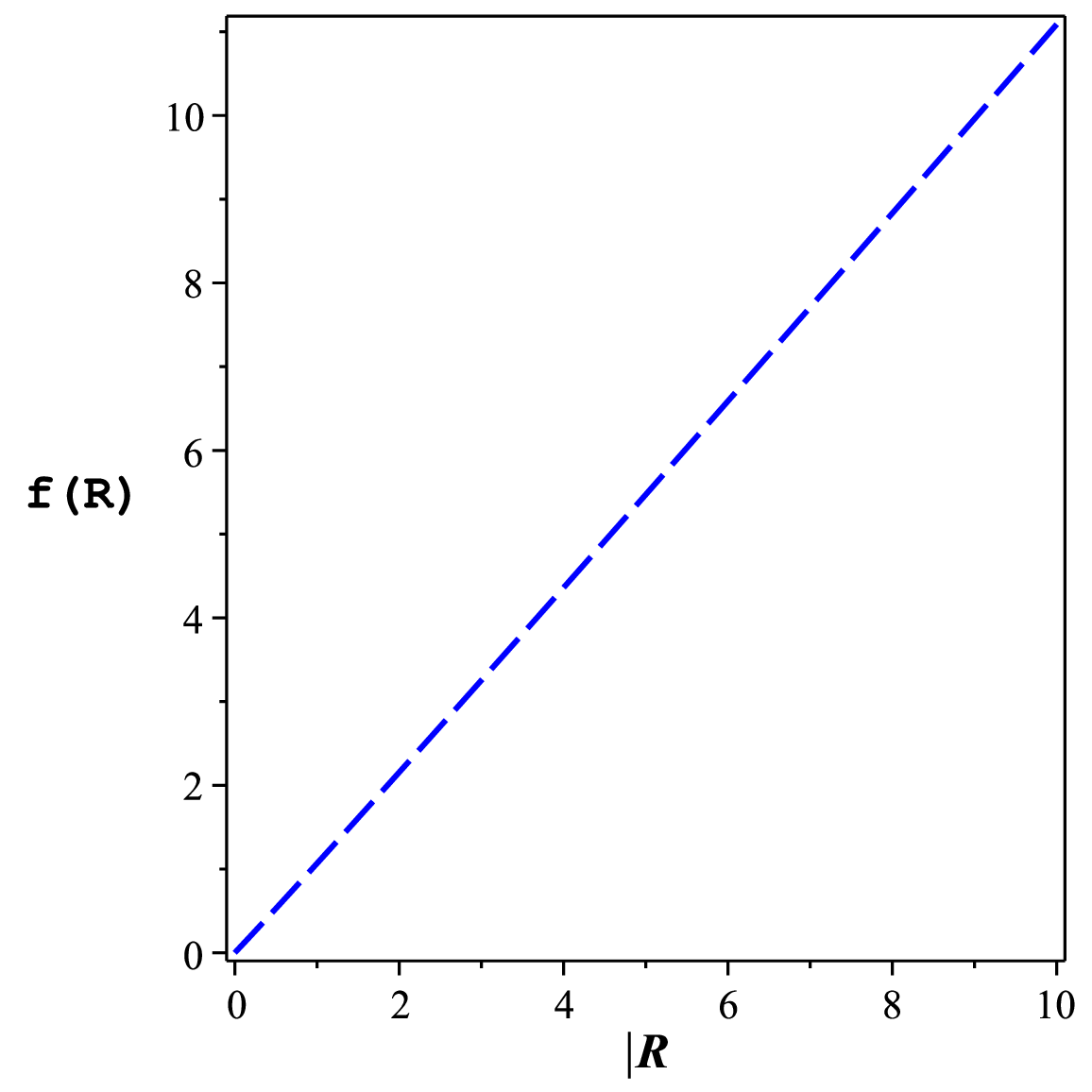}}\hspace{0.5cm}
\subfigure[~The behavior of the function $f_R$]{\label{fig:1d}\includegraphics[scale=0.25]{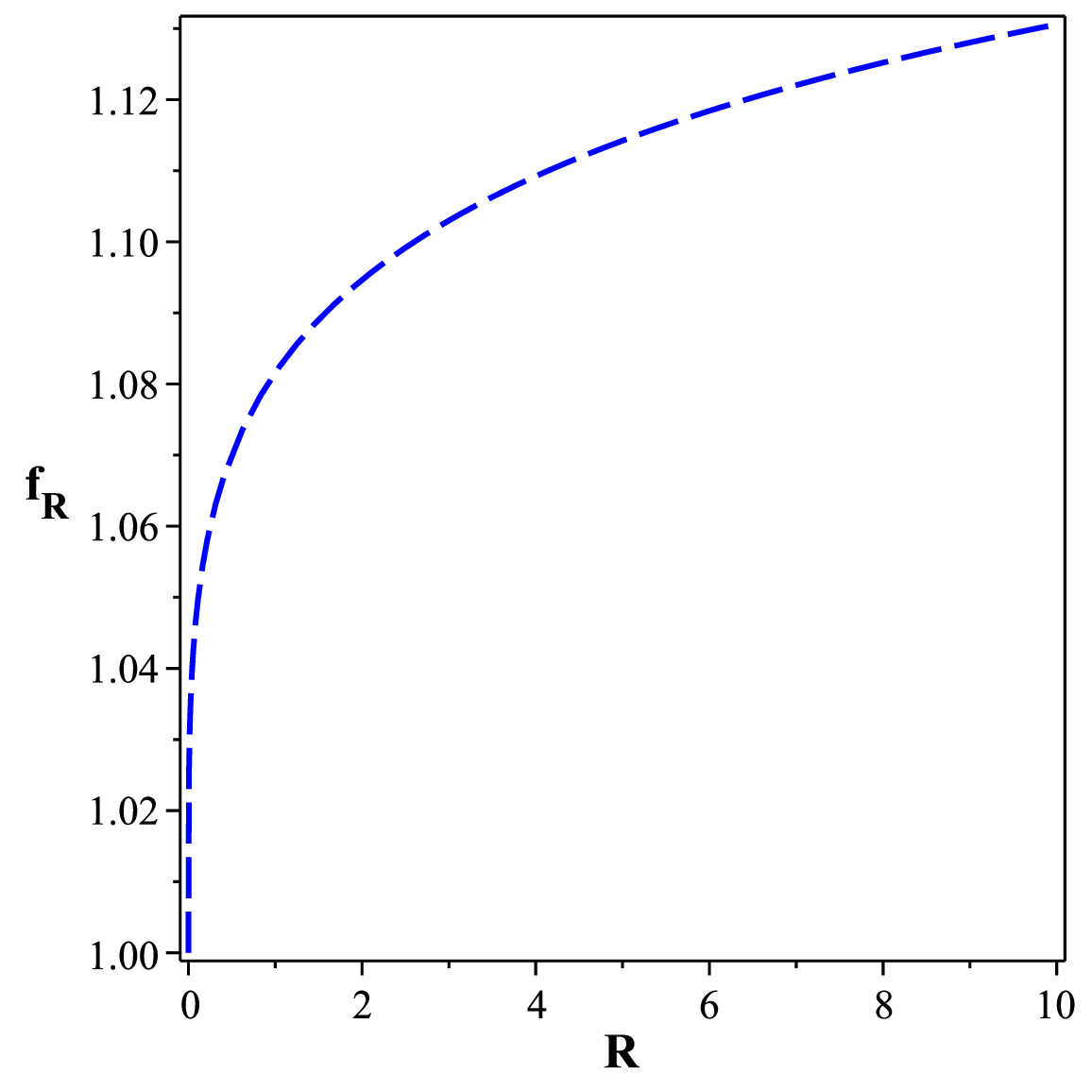}}\hspace{0.5cm}
\subfigure[~The behavior of the function $f_{RR}$]{\label{fig:1e}\includegraphics[scale=0.25]{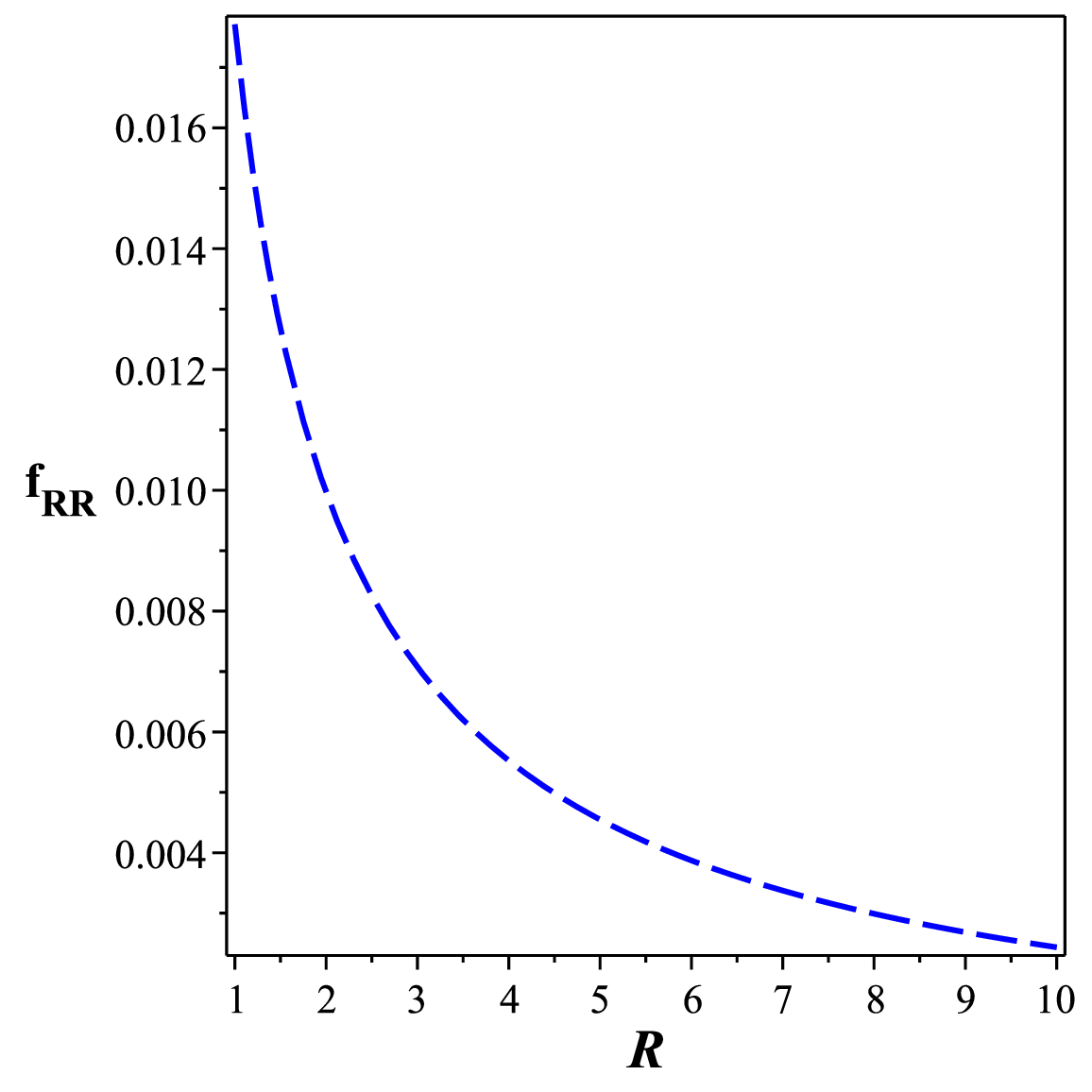}}\hspace{0.5cm}
\caption{Schematic plots of: Fig.~\subref{fig:1a} the behavior of the scalar function $\phi_1$;   Fig.~\subref{RS1}$\sim$ \subref{RS3} show the behavior of the physical singularity at the origin for $R$,  $R^{\mu\nu}R_{\mu\nu}$, and $R^{\mu\nu\alpha\beta}R_{\mu\nu\alpha\beta}$; Fig.~ \subref{fig:1b} shows the behavior of the potential $V$; finally, Fig.~\subref{fig:1c} $\sim$ \subref{fig:1e}  show the behavior of the functions $f(R)$, $f_R$ and $f_{RR}$.  Here we assume the following numerical values $M=1$, $\digamma=-100$, $c_2=-10$. }
\label{Figg:1}
\end{figure*}

Equation~\eqref{ST1} shows that all curvature invariants remain finite as $r \to \infty$, but diverge as
$r \to 0$, indicating the presence of a physical singularity at the origin, where we visually presented in Fig.~\ref{Figg:1} \subref{RS1}, \subref{RS2}, and \subref{RS3}, implies that the BH is not regular at this point. Thus, our solution of the BH has no impact on the emergence of a physical singularity at $r=0$. Compared to  GR, where the Schwarzschild solution yields the invariant set $\left\{R^{\mu\nu\alpha\beta}R_{\mu\nu\alpha\beta}, R^{\mu\nu}R_{\mu\nu}, R\equiv \left(\frac{1}{r^6}, 0,0\right) \right\}$  the singularities in our solution are generally stronger.
$r=2M$.

From the first equation in Eq.~\eqref{ST1}, we can derive the asymptotic form of $r(R)$ as,
\begin{align}\label{r(R)}
r(R)\approx\frac {\sqrt [4]{|2\digamma+m^2\|}}{\sqrt [4]{R}}\,.
\end{align}
Additionally, by substituting Eq.~(\ref{r(R)}) into the third equation of Eq.~(\ref{scal1}), we obtain,
\begin{eqnarray} \label{fR2}
&&f(R)\approx R+C_4R^{5/4}+C_5R^{3/2}+C_6R^{7/4}\,,
 \end{eqnarray}
 where $C_4$, $C_5$ and $C_6$ are constants that depend on $m$ and $\digamma$.
 Moreover, from Eq.~\eqref{scal1}, we find  that $r(\phi_1)$ is given by
 \begin{align}\label{sca}
 r(\phi_1)\approx-\frac{{e^{-256\,{\frac {\sqrt {3}c_1\,{\digamma}^{5/4}}{\sqrt {M} \sqrt {\|c_2\|} \left( 256\,\digamma-187\,{M}^{2} \right) }}}}}{\sqrt {\|c_2\|} \sqrt {M}\left( 256\digamma-187\,{c}^{2} \right)
} \left[ 256 \,\digamma\sqrt {c}\sqrt {-c_2}-187\,{c}^{5/2}\sqrt {-c_2}+256 \,\sqrt {3}{\phi_1}\,{\digamma}^{5/4} \right] \,.
 \end{align}
Equation~\eqref{sca} indicates that the constant of integration $c_2$
  must be negative, which is necessary to remove any ghosts from our solution. By combining Eqs.~\eqref{sca} and \eqref{scal1}, we obtain:
\begin{align}
V(\phi_1)\approx C_6+C_7\phi_1+C_8\phi_1{}^2\,,\label{Ve}
\end{align}
with $C_6$, $C_7$, and $C_8$ being constants that depend on  $m$, $c_2$ and $\digamma$.
The behavior of Eq.~\eqref{Ve} is illustrated in Fig.~\ref{Figg:1}\subref{fig:1b}, which shows a positive profile.

Next, we investigate whether the ansatz given by Eq.\eqref{sol} is physically acceptable within the framework of $f(R,G)$ gravity or not. To this end, we analyze the behavior of the functions $f(R)$, $f_R$, and $f_{RR}$, which are plotted in Fig.~\ref{Figg:1} \subref{fig:1c}, \subref{fig:1d}, and \subref{fig:1e}. These functions exhibit positive behavior, which is a necessary condition for a viable $f(R)$ theory, as discussed in Ref.~\cite{DeFelice:2010aj}.
\newpage
%%%%%%%%%%%%%%%%%%%%%%%%%%%%%%%%%%% Section 4 (Sec. V) %%%%%%%%%%%%%%%%%%%%%%%%%%%%%%%%%%%%%%%%
\section{Thermodynamics of the black hole}\label{S4}
%%%%%%%%%%%%%%%%%%%%%%%%%%%%%%%%%%%%%%%%%%%%%%%%%%%%%%%%%%%%%%%%%%%%%%%%%%%%%%%%%%%%%%
%%%%%%%%%%%%%%%%%%%%%%%%%%%%%%%%%%%%%%%%%%%%%%%%%%%%%%%%%%%%%%%%%%%%
%\subsection{Thermodynamics of the BH(\ref{elm1}) }\label{S5a}
%%%%%%%%%%%%%%%%%%%%%%%%%%%%%%%%%%%%%%%%%%%%%%%%%%%%%%%%%%%%%%%%%%%%%%%%%%%%%%%%%%%%%%%%%%%%%%%%%%
In this section, we thoroughly explore the thermodynamic aspects of our BH solution in $f(R,G)$ gravity.

\subsection{Thermodynamic properties of the black hole}
\begin{figure*}
\centering
\subfigure[~Metric behavior]{\label{fig:2a}\includegraphics[scale=0.25]{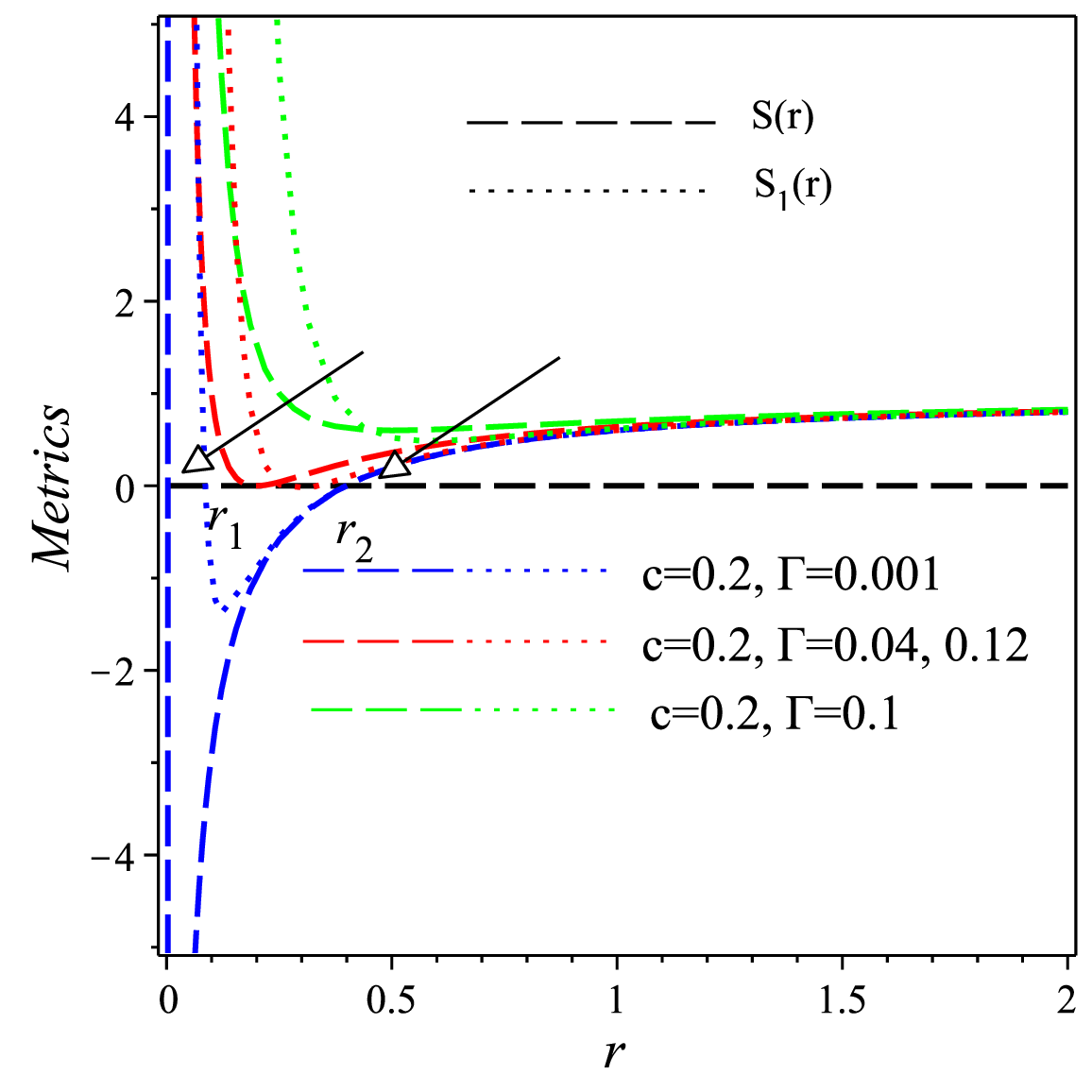}}\hspace{0.5cm}
\subfigure[~The entropy]{\label{fig:2b}\includegraphics[scale=0.25]{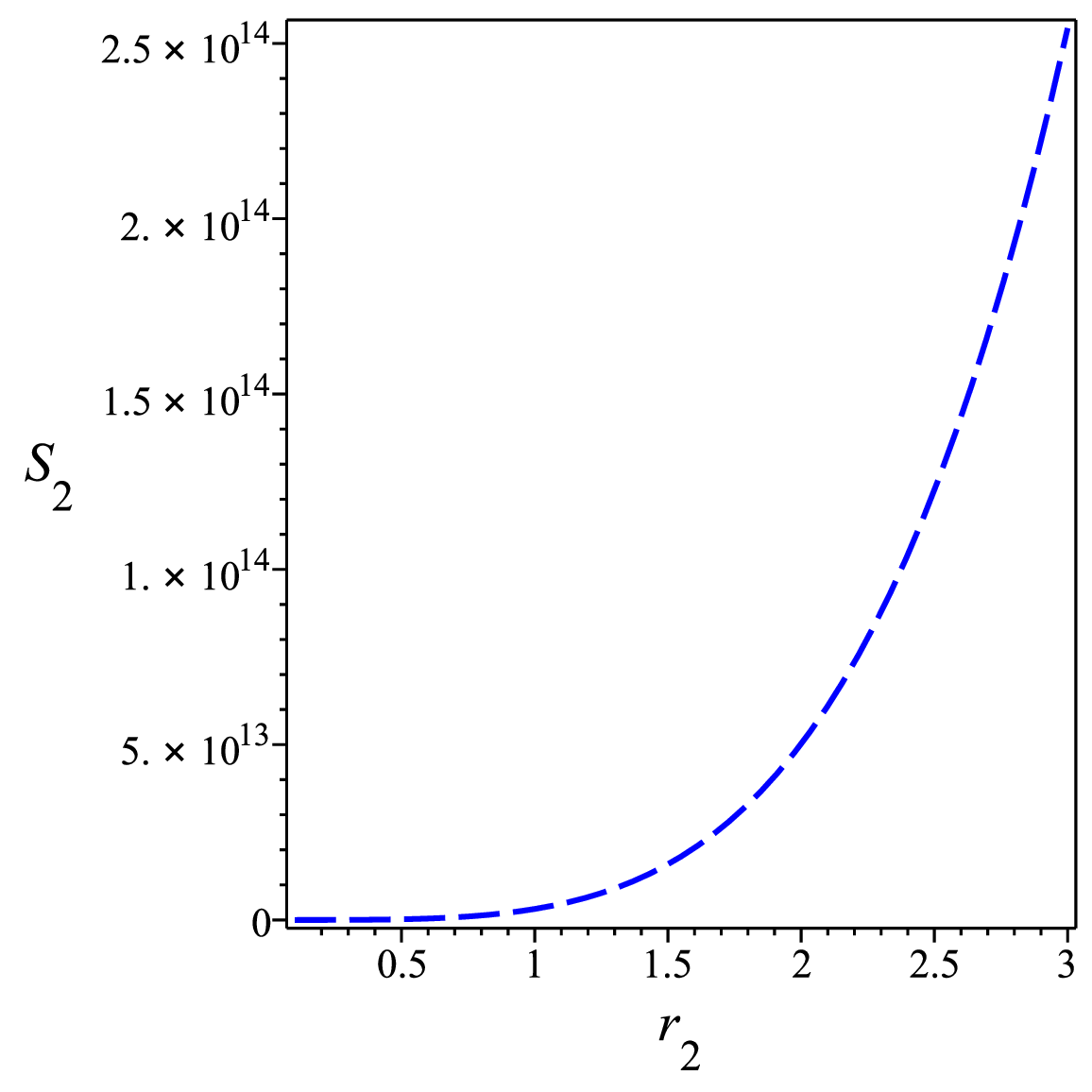}}\hspace{0.5cm}
\subfigure[~The  temperature]{\label{fig:2c}\includegraphics[scale=0.25]{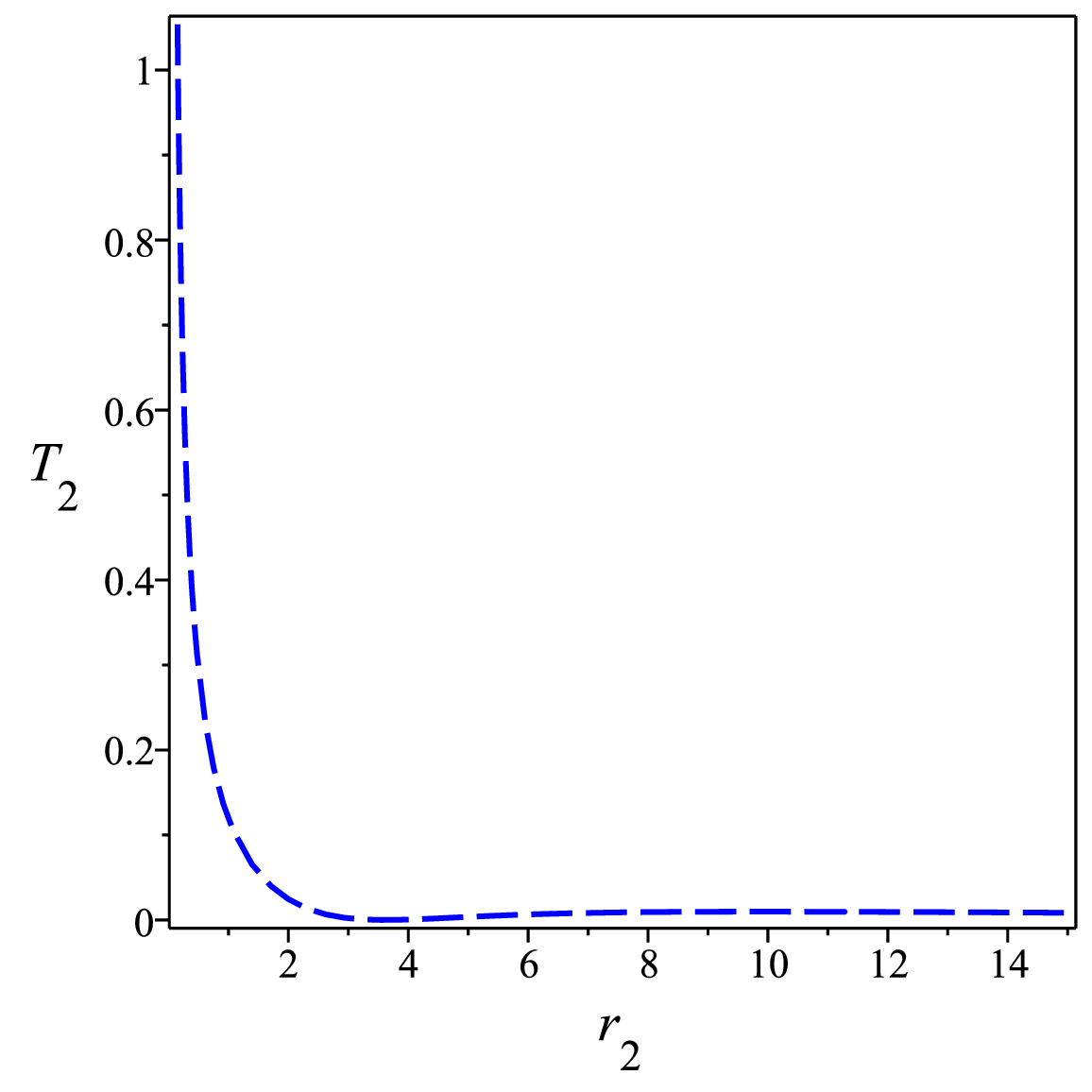}}\hspace{0.5cm}
\subfigure[~The heat capacity]{\label{fig:2d}\includegraphics[scale=0.25]{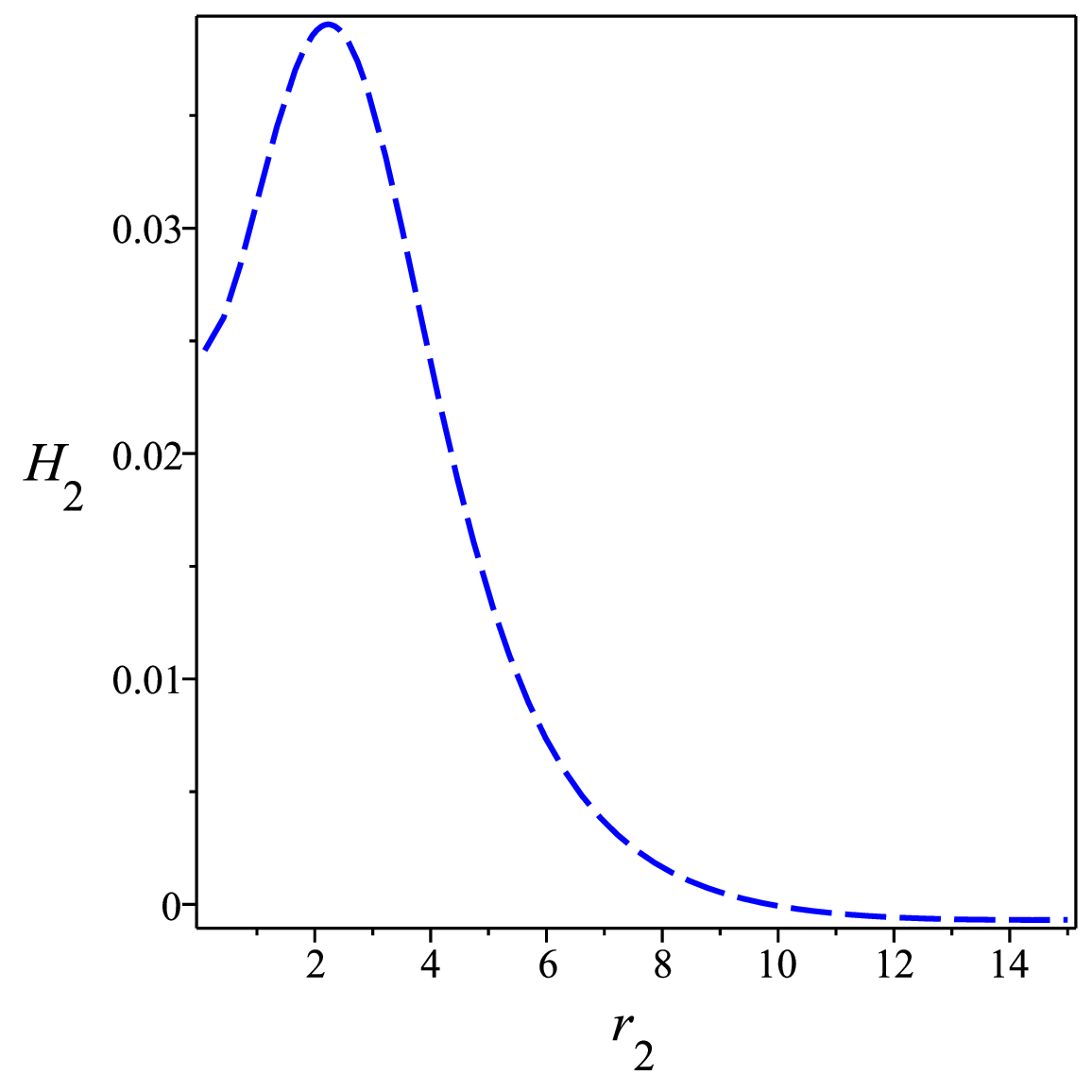}}\hspace{0.5cm}
\subfigure[~The   local energy]{\label{fig:2e}\includegraphics[scale=0.25]{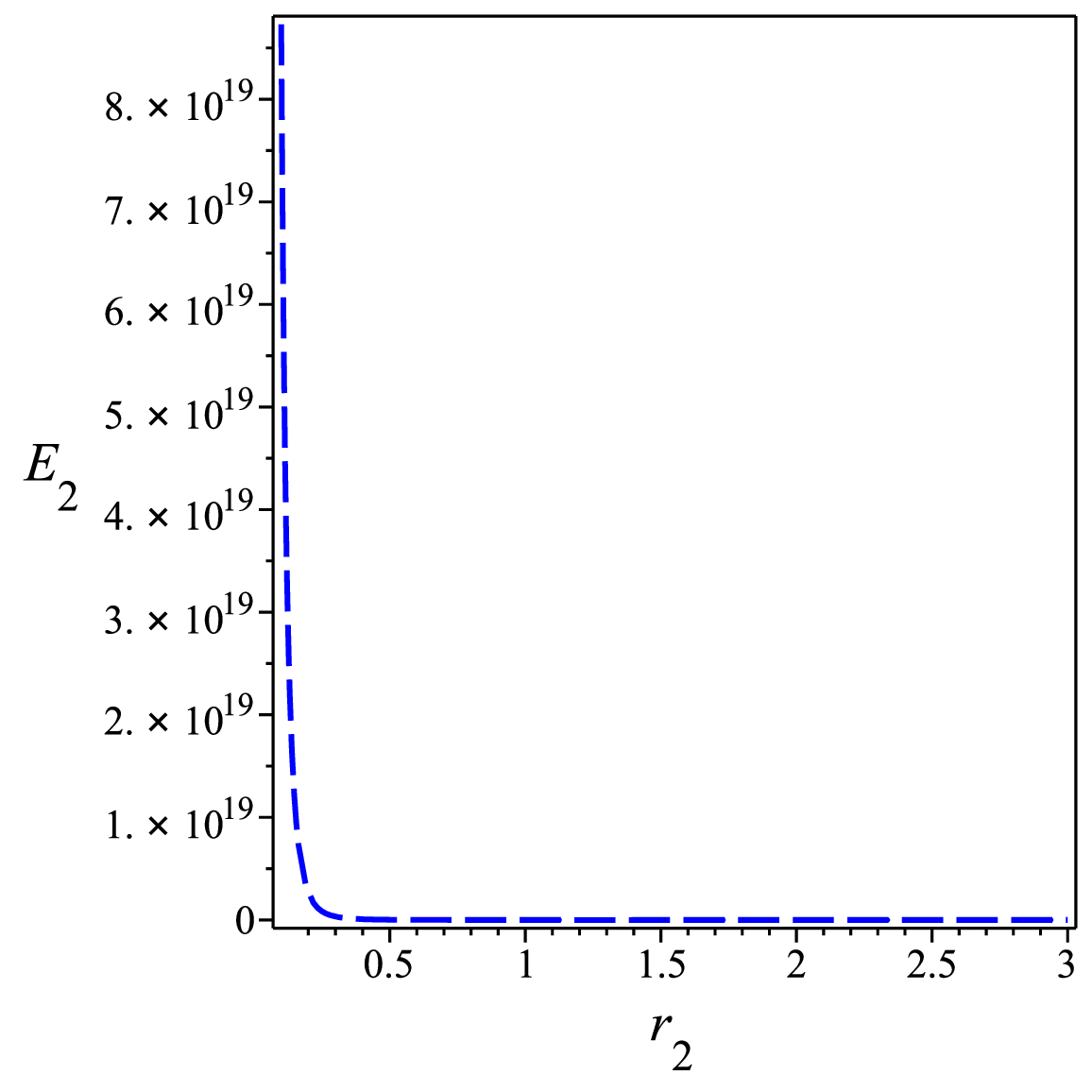}}\hspace{0.5cm}
\subfigure[~The horizon Gibbs free energy]{\label{fig:2f}\includegraphics[scale=0.25]{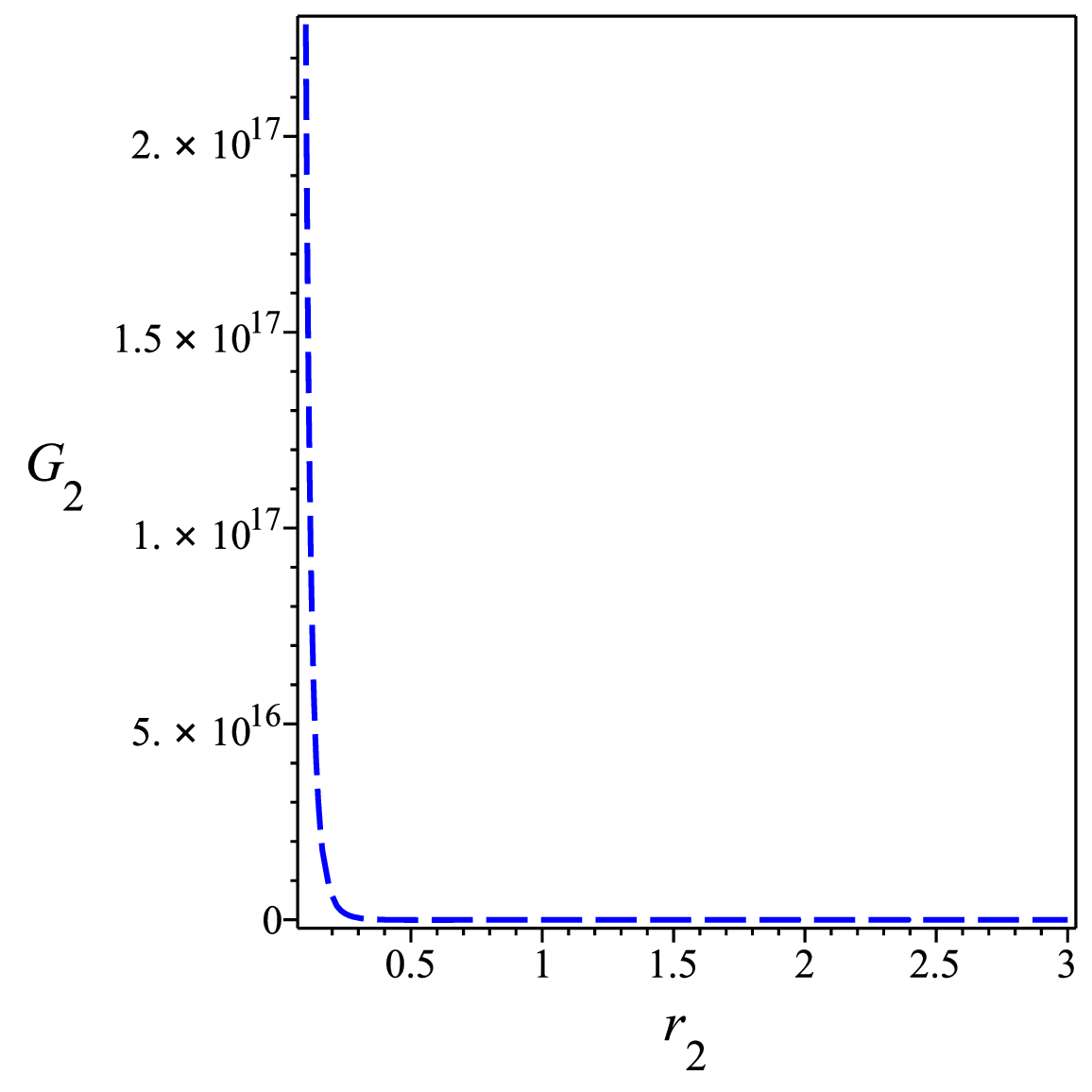}}
\caption{The behavior  the BH solution's thermodynamic quantities as given in Eq.~(\ref{sol}): \subref{fig:2a}
 the pattern  of the metric ansatz $S$ and $S_1$; \subref{fig:2b} the pattern   of the entropy; \subref{fig:2c} the behavior of the temperature and \subref{fig:2d} the behavior of the heat capacity of
 Eq.~(\ref{heat-cap1a11}); \subref{fig:2e} the pattern of the quasi-local energy, finally \subref{fig:2f} the pattern of the Gibbs free energy. The following numerical values are used in Fig.~\ref{Fig:2}, $\digamma=-100$, $c_1=1$, $c_2=-10$, $c_3=-100$, and $c_4=-100$. }
\label{Fig:2}
\end{figure*}

The temporal component of the metric potential in Eq.~(\ref{met3}) is expressed in the following manner
 \begin{eqnarray} \label{hor11}
&&S(r)=1-\frac{2M}{r}+\frac{\digamma}{r^2}\,. \end{eqnarray}
The pattern of the metric (\ref{met3}) is illustrated in Fig.~\ref{Fig:2}\subref{fig:2a}. This figure reveals that the  BH  described by this metric could have two distinct horizons.  The two horizons identified in the metric pattern of Eq.~(\ref{met3}) are $r_1$ and $r_2$. Specifically, $r_1$ denotes the inner Cauchy horizon of the BH, while $r_2$ represents the outer event horizon. The presence of these two horizons adds complexity to the BH's structure, with the inner Cauchy horizon marking the boundary beyond which the predictability of physical laws breaks down, and the outer event horizon representing the boundary from which nothing, not even light, can escape. This configuration is reminiscent of the structure in charged or rotating BHs in general relativity. In the case of $f(R)$ gravity or any modified gravity theory, observing such structures provides valuable insights into the theory's predictions, particularly concerning the spacetime geometry near compact objects. Understanding these horizons is crucial for comprehending the dynamics of BH interiors, the behavior of infalling matter, and the potential for observable phenomena such as echoes or ringdown signals following a merger event.  Figure~\ref{Fig:2}\subref{fig:2a} shows that for a parameter $\digamma=0.01$, the BH has two distinct horizons. However, when $\digamma$ is increased to $0.04$, the two horizons merge to form a single, degenerate horizon. This phenomenon occurs when $r_{1}=r_{2}=r_d$, indicating that the inner Cauchy horizon and the outer event horizon coincide at a single radial distance $r_d$. When the parameter $\digamma$ is increased further to $0.1$, the analysis reveals that we enter a parameter regime where the BH does not possess any horizons.

By applying the condition $S(r_2) = 0$, the total mass inside the event horizon ($r_2$) can be determined. Subsequently, one can obtain the mass-radius relation for the horizon, given by
\begin{eqnarray} \label{hor-mass-rad1a11}
&& {r_2}=M+\sqrt{M^2-\digamma}\, .
\end{eqnarray}
Equation \eqref{hor-mass-rad1a11} shows that the horizon is different from the Schwarzschild one due to the effect of $\digamma$.
Equation \eqref{hor-mass-rad1a11} shows that as $\digamma=0$ we get the horizon of the Schwarzschild solution.
The entropy in $f(R,G)$ theory is represented as \cite{PhysRevD.84.023515,Zheng:2018fyn}
\begin{equation}\label{ent11}
S(r_2)=\frac{1}{4}Af_R,
\end{equation}
where $A$ is the area and
$f_R$ denotes the first-order derivative of the function
$f(R)$ taken with respect to the Ricci scalar. As Eq. \eqref{ent11} shows that the entropy has been affected by the derivative of $f(R)$.  By substituting the value of $f_R=F(r)$\footnote{We did not list the value of $f_R$ due to its lengthy but we present its asymptote as given by Eq.\eqref{scal1}.} in Eq.~(\ref{ent11}), the entropy of the BH described by Eq.~(\ref{met3}) is given by
\begin{align} \label{ent1}
&{S_2}_{{}_{{}_{{}_{{}_{\tiny \mathrm{Eq.}~(\ref{met3})}}}}}=\frac {92223 \sqrt {2}\pi}{32768{\digamma}^{6}{r_2}^{5/2}\sqrt {\digamma+{r_2}^{2}}}  \left[ {\frac {16384}{92223}}
c_3{\digamma}^{6}\sqrt {2}{r_2}^{13/2}\sqrt {\digamma+{r_2}^{2}}+
c_2 \left( -{\frac {8192}{92223}}{\digamma}^{3}{r_2}^{5}
 \left( {\digamma}^{2}-6\digamma{r_2}^{2}+{r_2}^{4} \right) \ln  \left(
r_2 \right) +{\digamma}^{7}\right.\right.\nonumber\\
&\left.\left.-{\frac {43264}{92223}}{r_2}^{2}
 \left( r_2-{\frac {54865}{43264}} \right) {\digamma}^{6}-{\frac {8192}{
92223}}{r_2}^{4} \left( {\frac {44597}{8192}}+{r_2}^{2}+{
\frac {267}{80}}r_2 \right) {\digamma}^{5}-{\frac {8192}{92223}}{r_2}^{6} \left( {r_2}^{2}+{\frac {8235}{8192}}-{\frac {943}{
240}}r_2 \right) {\digamma}^{4}\right.\right.\nonumber\\
&\left.\left.-{\frac {65024}{461115}}{r_2}^{8
} \left( {\frac {61295}{65024}}+r_2 \right) {\digamma}^{3}+{\frac {256}{
30741}}{r_2}^{10} \left( -{\frac {9805}{768}}+r_2 \right)
{\digamma}^{2}+{\frac {1433}{92223}}\digamma{r_2}^{12}-{\frac {25}{92223}}{
r_2}^{14} \right)  \right] \,.
\end{align}
Equation \eqref{ent1} indicates that the expression of entropy is affected by the higher-order curvature terms. The entropy function governed by Eq.~(\ref{ent1}) is plotted in Fig.~\ref{Fig:2} \subref{fig:2b}, demonstrating its positive nature for all the values of $r_{2}$.

By adopting the procedure as described in Refs.~\cite{PhysRevD.86.024013,Sheykhi:2010zz,Hendi:2010gq,PhysRevD.81.084040}, we determine the Hawking temperature, which is defined as
  \begin{equation}\label{temp11}
T_2 = \frac{\chi}{2\pi}=\frac{1}{4\pi}\left[\frac{1}{\sqrt{-g_{tt}g_{rr}}}\frac{dg_{tt}}{dr}\right]_{r\to r_2},
\end{equation}
where $\chi$ is the surface gravity of the BH. By using Eqs.~\eqref{met3} and (\ref{temp11}), we find
%that the Hawking temperature %is represented as
\begin{eqnarray}\label{T11}
&&T_2=\frac { \left( {r_2}^{2}+\digamma \right) ^{3/2}}{2\pi {r_2} \sqrt {{
{ \left( {r_2}^{2}-\digamma \right) ^{2} \left(8\digamma{r_2}^{3} -{\digamma}^{3}-3{\digamma}^{2}{r_2}^{2}-3\digamma{r_2}^{4}-{r_2}^{6}
 \right) }}}} \,.
\end{eqnarray}
The pattern of  Eq.~(\ref{T11}) is shown in Fig.~\ref{Fig:2} \subref{fig:2c} which indicates that $T_2$ is always positive.

Additionally, it is possible to analyze the stability of the BH solution at both the dynamic and perturbative levels~\cite{Nashed:2003ee,Myung:2011we,Myung:2013oca}.  To examine the thermodynamic stability of BHs,
the heat capacity  $H(r_2)$ near the event horizon is calculated as \cite{Nouicer:2007pu,DK11,Chamblin:1999tk}
\begin{equation}\label{heat-capacity11}
H_2\equiv H(r_2)=\frac{\partial r_2}{\partial T_2}=\frac{\partial r_2}{\partial r_2} \left(\frac{\partial T_2}{\partial r_2}\right)^{-1}\, .
\end{equation}
If the BH's heat capacity $H_2$ is positive, it will be thermodynamically stable.  However, if $H_2$ is negative, it will be unstable.
 By replacing Eq.~(\ref{hor-mass-rad1a11}) and (\ref{T11}) with Eq.~(\ref{heat-capacity11}), the heat capacity can be represented as
\begin{align}\label{heat-cap1a11}
&{H_2}_{{}_{{}_{{}_{{}_{\tiny \mathrm{Eq.}~(\ref{met3})}}}}}=\frac{\left( {\digamma}^{5}+11\,{\digamma}^{4}{r_2}^{2}-20\,{r_2}^{3}{\digamma}^{3}+26
\,{\digamma}^{3}{r_2}^{4}-40\,{r_2}^{5}{\digamma}^{2}+22\,{\digamma}^{2}{r_2
}^{6}+5\,\digamma{r_2}^{8}-4\,{r_2}^{7}\digamma-{r_2}^{10} \right)}{\pi\left\{ {\digamma}^{3}+3\,{\digamma}^{
2}{r_2}^{2}+ \left( 3\,{r_2}^{4} -8\,{r_2}^{3}\right)
\digamma+{r_2}^{6} \right\} ^{3/2}}\, .
\end{align}
Equation (\ref{heat-cap1a11}) shows that $H_2$ does not locally diverge.
The heat capacity is illustrated in Fig.~\ref{Fig:2} \subref{fig:2d},  which demonstrates that it has a positive pattern, which means that the BH given by Eq.~(\ref{met3}) is stable.

 The  local energy of the theory under consideration takes the following form  \cite{PhysRevD.84.023515,PhysRevD.86.024013,Sheykhi:2010zz,Hendi:2010gq,PhysRevD.81.084040,Zheng:2018fyn}
\begin{equation}\label{en}
E(r_2)=\frac{1}{4}\displaystyle{\int }\Bigg[2f_{R}(r_2)+r_2{}^2\Big\{f(R(r_2))-R(r_2)f_{R}(r_2)\Big\}\Bigg]dr_2.
\end{equation}
By combining Eqs. (\ref{met3}) and (\ref{en}), we find the quasi-local energy, which is lengthy.
The quasi-local energy behavior is shown in Fig.~\ref{Fig:2}\subref{fig:2e}, which has a positive pattern.  The Gibbs free energy is given by \cite{Zheng:2018fyn,Kim:2012cma}
\begin{equation} \label{enr11}
\mathbb{G}(r_2)=M(r_2)-T(r_2)S(r_2)\,.%+P(r_+)V(r_+),
\end{equation}
The quantities $M(r_2)$, $T(r_2)$, and $S(r_2)$ are the mass, temperature, and entropy at the event horizon, respectively. By inserting Eqs.~(\ref{hor-mass-rad1a11}),~(\ref{ent1}), and (\ref{T11}) into Eq.~(\ref{enr11}), we compute the Gibbs free energy, which is very lengthy. We depict the Gibbs free energy of the BH described by the metric in Eq.~(\ref{met3})  in Fig.~\ref{Fig:2} \subref{fig:2f}, which indicates that the Gibbs energy is always positive.

Let us summarize our findings: we have analyzed the thermodynamic properties of the black hole solution within the framework of modified
$f(R)$ gravity. Special attention has been given to how each thermodynamic quantity, such as temperature, entropy, and Gibbs free energy, is affected by the modifications to the gravitational action. The deviations from the standard General Relativity results have been qualitatively highlighted and discussed. To further illustrate these effects, the relevant plots in Fig.~\ref{Fig:2}\subref{fig:2d}, \subref{fig:2e}, and \subref{fig:2f} depict the physical behavior of these thermodynamic quantities and support the analytical results. These findings provide deeper insight into the thermodynamic stability and physical viability of the black hole solution in the context of modified gravity.

\subsection{Thermodynamic topology via Barrow entropy}
This subsection explores the thermodynamic topology of the BH solution in $f(R,G)$ gravity by using Barrow entropy. Our motive for using Barrow entropy is to incorporate the quantum gravitational effect, which certainly impacts the relationship between the surface area of the horizon and entropy. For example, if we consider that the horizon of the BH has a fractal-like structure, then its area of the horizon will also be increased, which also impacts the entropy of the BH to $r^{1+\delta/2}$. Firstly, we describe its basic framework  suggested by Barrow entropy~\cite{Barrow:2020tzx} (for more details regarding the impact of Barrow entropy in the BH thermodynamics, check Ref.~\cite{Capozziello:2025axh}), and it can be written as
%in the terms of $f(R,G)$
%in the given form
\begin{eqnarray}\label{Br1}
S_\mathrm{B}=\left(\frac{A f_{R}}{A_{\mathrm{Pl}}}\right)^{1+\delta/2}=(\pi r^{2} f_{R})^{1+\delta/2},
\end{eqnarray}
where \( A_\mathrm{Pl} \) represents Planck's area. The term \( f_R \) in Eq.~(\ref{Br1}) originates from the contribution of \( f(R,G) \) gravity, as described in Eq.~(\ref{ent11}). Since we are working within modified gravity, we depart from GR, meaning \( f_R \) cannot be arbitrarily chosen. Notably, \( f_R \) depends on the horizon radius \( r_2 \), requiring numerical evaluation. To determine \( f_R \), we differentiate the first two terms of Eq.~\eqref{fR2}, then substitute Ricci scalar to incorporate the effects of \( f(R, G) \) gravity. Our goal is to obtain \( f_R = 1.48765 \) numerically, simplifying our analysis of the topological interpretation.

One might ask why we do not simply set \( f_R = 1 \). The reason is that this would correspond to GR, whereas our BH solution exists in the modified \( f(R, G) \) framework. The numerical value of \( f_R \) is determined using the fixed-point method, and varying this value may affect the location of the divergence point.

We discuss the thermodynamic topology of our BH solution by deriving the mass of the BH in the form of Barrow entropy.  Furthermore, we mentioned here that if we put the parameter $\delta=0$, we can get the Bekenstein-Hawking entropy as mentioned in Eq.~(\ref{ent11}), while by substituting $\delta=1$, we can obtain the most fractal structure.

%begin our study by discussing the  We thermodynamic topology of this solution by employing  Eq.~(\ref{hor-mass-rad1a11}) and (\ref{Br1}), one can get the mass in terms of Barrow entropy, which is given as
\begin{eqnarray}\label{Br2}
M_{2}(S_\mathrm{B},~\digamma)=\frac{\left(\pi ^{-\frac{\delta }{2}-1} S_\mathrm{B}\right)^{-\frac{1}{\delta +2}} \left\{f_{R} \ \digamma+\left(\pi ^{-\frac{\delta }{2}-1} S_\mathrm{B}\right)^{\frac{2}{\delta +2}}\right\}}{2 \sqrt{f_{R}}}.
\end{eqnarray}
We can determine the conjugate temperature corresponding to  Barrow entropy from Eq.~(\ref{Br2}) by using the relation $\partial{M_{2}}/\partial{S_\mathrm{B}}$, which can be written as
\begin{eqnarray}\label{Br3}
T_\mathrm{B}(S_\mathrm{B},~\digamma)=\frac{\left(\pi ^{-\frac{\delta }{2}-1} S_\mathrm{B}\right)^{-\frac{1}{\delta +2}} \left\{\left(\pi ^{-\frac{\delta }{2}-1} S_\mathrm{B}\right)^{\frac{2}{\delta +2}}-\digamma f_R\right\}}{2 (\delta +2) S_\mathrm{B} \sqrt{f_R}}.
\end{eqnarray}
Furthermore, to determine the heat capacity in the form of Barrow entropy, we adopted the same methodology as described in Ref.~\cite{Bhattacharya:2024bjp} by using Eq.~(\ref{Br3}). As a result, we have
\begin{eqnarray}\label{Br4}
C_\mathrm{B}(S_\mathrm{B},~\digamma)=\left\{\frac{1}{\frac{\partial{T_\mathrm{B}(S_\mathrm{B},~\digamma)}}{\partial{S_\mathrm{B}}}}\right\}\frac{1}{T_\mathrm{B}(S_\mathrm{B},~\digamma)}=\frac{(\delta +2) S_\mathrm{B} \left\{\left(\pi ^{-\frac{\delta }{2}-1} S_\mathrm{B}\right)^{\frac{2}{\delta +2}}-f_{R} \digamma\right\}}{(\delta +3) f_{R} \digamma-(\delta +1) \left(\pi ^{-\frac{\delta }{2}-1} S_\mathrm{B}\right)^{\frac{2}{\delta +2}}}.
\end{eqnarray}
If we look at Eq.~(\ref{Br4}), we can determine the zero points or divergence points from the denominator, which are for $\delta=0,~0.5,~1$ are $1.40208,~1.11438$ and $0.903693$, respectively. Here, we mentioned that we have fixed the $\digamma=0.1$ and $f_{R}=1.48765$ to compute these zero points.

Recently, researchers have introduced a new approach that allows BH properties to be distinguished without depending on the spacetime's dynamic equations or structure analysis. This method was previously adopted within a coordinate space of BHs to determine the location of light rings by employing the null geodesics \cite{Cunha:2017qtt,Cunha:2020azh}. It has been demonstrated that light rings are positioned at the zeros of a particular field formed within a coordinate framework of the BH spacetime. It was observed that asymptotically flat spacetimes with spherical topology of killing horizon, BHs were found to illustrate at least one external light ring for each rotational orientation \cite{Cunha:2020azh}, and this concept was extended to other scenarios \cite{Guo:2020qwk,Wei:2020rbh}. This topological approach was subsequently applied to various areas of the BH thermodynamics as discussed in Refs.~\cite{Wei:2022dzw,Wei:2021vdx,Yerra:2022alz,Yerra:2022eov,Gogoi:2023qku,Gogoi:2023xzy,Fan:2022bsq,Fang:2022rsb,Yerra:2022coh}. It has been established that the critical point in the BH phase transition may correspond to the non-zero topological charge, offering a topological viewpoint \cite{Wei:2021vdx,Yerra:2022alz,Gogoi:2023qku,Gogoi:2023xzy}. Researchers have applied this topological interpretation to investigate different types of phase transition, such as Van der Waals phase transition \cite{Wei:2021vdx,Yerra:2022alz,Gogoi:2023qku,Gogoi:2023xzy}, Hawking-Page phase transition \cite{Fang:2022rsb,Yerra:2022coh}, extremal and Davis-type phase transition \cite{Bhattacharya:2024bjp,Zafar:2025sxl}, by using Duan's topological $\phi$-mapping theory \cite{Duan:1979ucg,Duan:1984ws}. The fundamental concept in this approach is to construct a vector field (usually by using temperature and free energies) and examine its winding number around critical points. This winding number provides us with a topological charge (commonly an integer) that remains unchanged under any small changes that occur in the system, ensuring its stability against small fluctuations. In our analysis, we adopted this methodology by constructing a vector field using the temperature and examining the topological charge for our BH solution in $f(R,G)$ gravity. The winding number associated with this vector field offers a framework to characterize the phase transitions and establishes a correspondence between the geometry of the phase space of thermodynamics and physical features like the presence of critical points and the stability of BHs in $f(R,G)$ gravity. In addition to the conventional thermodynamic method, this approach incorporates the geometric interpretation to better understand the phase structure of BHs in $f(R,G)$ gravity. The basic framework for thermodynamic topology is given in detail in Refs.~\cite{Bhattacharya:2024bjp,Wei:2020rbh,Wei:2022dzw,Wei:2021vdx,Yerra:2022alz,Yerra:2022eov,Gogoi:2023xzy,Gogoi:2023qku,Yerra:2022coh,Wei:2022mzv}. Here, we describe the thermodynamic topology of this BH solution directly by defining a thermodynamic potential, and it is given as
\begin{eqnarray}\label{TP1}
\Phi_\mathrm{B}(S_\mathrm{B},~\theta)=\frac{1}{\sin\!\theta \ T_\mathrm{B}(S_\mathrm{B})}=\frac{2 (\delta +2) S_\mathrm{B} \sqrt{f_R} \left(\pi ^{-\frac{\delta }{2}-1} S_\mathrm{B}\right)^{\frac{1}{\delta +2}}}{\sin\!\theta\left\{\left(\pi ^{-\frac{\delta }{2}-1} S_\mathrm{B}\right)^{\frac{2}{\delta +2}}-\digamma f_R\right\}}.
\end{eqnarray}
The incorporation of the term $1/\sin\!\theta$ is due to the simplification of this analysis as detailed in Refs.~\cite{Bhattacharya:2024bjp,Wei:2020rbh,Wei:2022dzw,Wei:2021vdx,Yerra:2022alz,Yerra:2022eov,Gogoi:2023xzy,Gogoi:2023qku,Yerra:2022coh,Wei:2022mzv}. Furthermore, we can derive the vector field by utilizing Eq.~(\ref{TP1}), that can be expressed as
\begin{eqnarray}\label{TP2}
\phi^{S_\mathrm{B}}&=&\partial_{S_\mathrm{B}}\Phi(S_\mathrm{B},~\theta)=\frac{2 \sqrt{f_{R}} \csc\!\theta  \left(\pi ^{-\frac{\delta }{2}-1} S_\mathrm{B}\right)^{\frac{1}{\delta +2}} \left\{(\delta +1) \left(\pi ^{-\frac{\delta }{2}-1} S_\mathrm{B}\right)^{\frac{2}{\delta +2}}-(\delta +3) f_{R} \ \digamma\right\}}{\left\{\left(\pi ^{-\frac{\delta }{2}-1} S_\mathrm{B}\right)^{\frac{2}{\delta +2}}-f_{R} \ \digamma\right\}^2},\\\label{TP3}
\phi^\theta&=&\partial_{\theta}\Phi(S_\mathrm{B},~\theta)=-\frac{2 (\delta +2) \sqrt{f_{R}} S_\mathrm{B} \cot\!\theta \csc\!\theta \left(\pi ^{-\frac{\delta }{2}-1} S_\mathrm{B}\right)^{\frac{1}{\delta +2}}}{\left(\pi ^{-\frac{\delta }{2}-1} S_\mathrm{B}\right)^{\frac{2}{\delta +2}}-f_{R} \ \digamma}.
\end{eqnarray}
Here, we can notice the significance of the term $1/\sin\!\theta$, which is useful in this study to obtain the zero point. The range for $\theta \in [0,~\pi]$ because at $\theta=\pi/2$, $\phi^{\theta}$ becomes zero which showed the zero point. It is easy to determine the topological current by using Duan's topological current theory, as detailed in Refs.~\cite{Duan:1979ucg, Duan:1984ws}, and it takes the given form
\begin{eqnarray}\label{TC1}
\mathfrak{j}^{\mathfrak{u}}=\partial_{\mathfrak{v}}V^{\mathfrak{u}\mathfrak{v}}=\frac{1}{2\pi}\varepsilon^{\mathfrak{u}\mathfrak{v}\mathfrak{p}}\varepsilon_{\mathfrak{a}\mathfrak{b}}\partial_{\mathfrak{v}}n^{\mathfrak{c}}\partial_{\mathfrak{p}}n^{\mathfrak{d}},
\end{eqnarray}
where $\mathfrak{u},~\mathfrak{v},~\mathfrak{p}=0,~1,~2$ and the normalized vector is expressed as $n^{\mathfrak{c}}=(\frac{\phi^{S_\mathrm{B}}}{||\phi||},~\frac{\phi^\theta}{||\phi||})$. Here, $V^{\mathfrak{u}\mathfrak{v}}$ is an anti-symmetric super-potential. Moreover, the conservation of the resulting topological currents is ensured by Noether's theorem $\partial_{\mathfrak{u}}\mathfrak{j}^{\mathfrak{u}}=0$. Now, in order to get the topological number, we need to modify this current by employing the following expression
\begin{eqnarray}\label{TC2}
\mathfrak{j}^{\mathfrak{u}}=\delta^{2}(\phi) J^{\mathfrak{u}}\left(\frac{\phi}{\mathfrak{x}}\right).
\end{eqnarray}
One can define the Jacobi tensor, which is written as
\begin{eqnarray}\label{TC3}
\varepsilon^{\mathfrak{c}\mathfrak{d}}J^{\mathfrak{u}}\left(\frac{\phi}{\mathfrak{x}}\right)=\varepsilon^{\mathfrak{u}\mathfrak{vu}\mathfrak{p}}\partial_{\mathfrak{v}}\phi^{\mathfrak{c}} \partial_{\mathfrak{p}}\phi^{\mathfrak{d}}.
\end{eqnarray}
When $\mathfrak{u}$ is set to zero, the Jacobi vector takes on the conventional Jacobi form, which is clearly reflected in the left second term of Eq.~(\ref{TC3}). It is evident from the equation that $\mathfrak{j}^{\mathfrak{u}}$ remains nonzero solely when $\phi$ is zero, and we then obtain the total charge $
W$ through further algebraic steps, which yields
\begin{eqnarray}\label{TC4}
W=\int_{\Sigma}\mathfrak{j}^{0}d^2 \mathfrak{x}=\sum_{i=1}^{\mathfrak{n}} \eta_{\mathfrak{i}} \ \beta_{\mathfrak{i}}=\sum_{\mathfrak{i}=1}^{\mathfrak{n}}\omega_{\mathfrak{i}}=\frac{1}{2 \ \pi} \oint_{\mathcal{C}_{\mathfrak{i}}}d\Omega.
\end{eqnarray}
The term $\beta_{\mathfrak{i}}$ refers to the positive Hopf index, which is used to calculate the number of loops the vector $\phi^{\mathfrak{a}}$ forms within the $\phi$ space around the zero point $z_{\mathfrak{i}}$. Additionally, $\eta_{\mathfrak{i}}$ specifies whether the orientation at the zero point $z_{\mathfrak{i}}$ is positive or negative, indicated by $\pm 1$, within the topological context. The winding number ${\omega}_{i}$, linked with the $\mathfrak{i}$-th zero point of $\phi$ in the region $\Sigma$, remains constant and is unaffected by any changes in the geometry of that region. Furthermore, assuming that $C_{i}$ is a positively oriented circle enclosing the $\mathfrak{i}$-th zero point, the associated winding number can be readily determined and expressed as follows
\begin{eqnarray}\label{JBT}
\omega_{i}=\frac{1}{2 \ \pi} \oint_{\mathcal{C}_{i}}d\Omega.
\end{eqnarray}

\begin{figure*}[ht]
\centering
\subfigure[~$\delta=0$]{\label{delta=0}\includegraphics[scale=0.5]{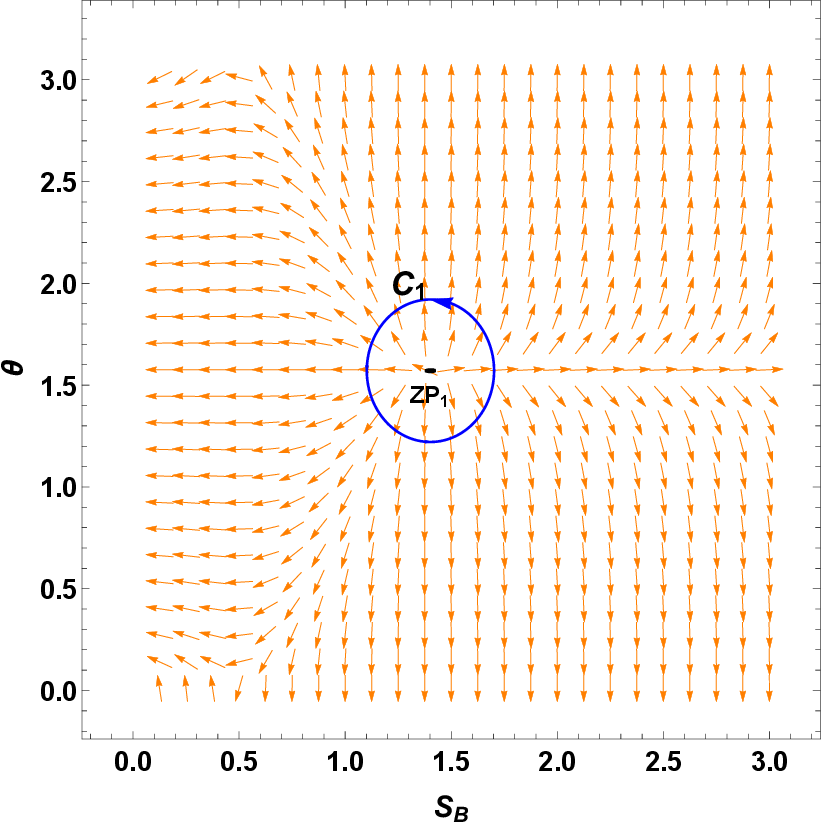}}\hspace{0.5cm}
\subfigure[~$\delta=0.5$]{\label{delta=0.5}\includegraphics[scale=0.5]{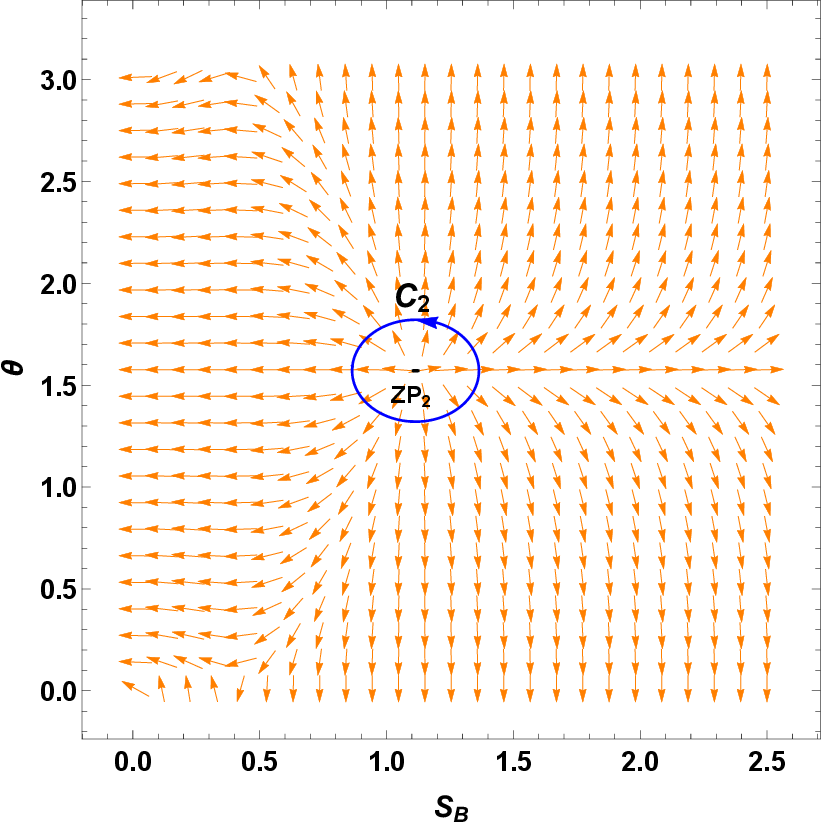}}\hspace{0.5cm}
\subfigure[~$\delta=1$]{\label{delta=1}\includegraphics[scale=0.5]{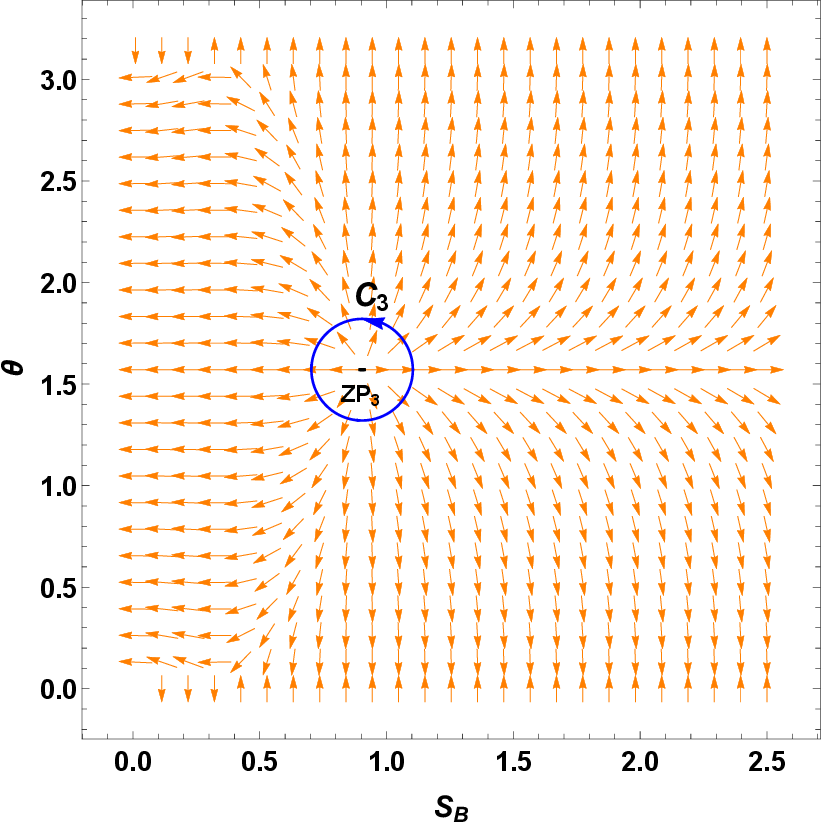}}\hspace{0.5cm}
\caption{Schematic plots of a normalized vector field in $S_\mathrm{B}-\theta$ plan. In Fig.~\subref{delta=0}, we have presented the zero point or divergence points $ZP_{1},~ZP_{2},~ZP_{3}$ with the black dots, which are surrounded by the circles in blue color $C_{1},~C_{2},~C_{3}$ for $\delta=0$ in Fig.~\subref{delta=0}, $\delta=0.5$ in Fig.~\subref{delta=0.5} and $\delta=1$ in \subref{delta=1}, respectively. We plot these normalized vectors by varying the parameter $\delta$; we have fixed $f_{R}=1.48765$ and $\digamma=0.1$. }
\label{TTTT}
\end{figure*}
The above-mentioned formalism yields a topological charge (winding number), which serves as a consistent and geometric indicator for distinguished BHs' phase structure. Furthermore, non-zero values of topological charge, such as $\pm1$, indicate the critical points that correspond to a phase transition, reflecting that BHs transit from one thermodynamic phase to another (stable to unstable configuration or vice versa). For example, a winding number of $1$ reflects the stable configuration of BH, while $-1$ points to an unstable phase, while topological charge $0$ reflects the existence of both stable and unstable configurations (or it also indicates the absence of phase transition) in the thermodynamic structure.
Moreover, we can compute the winding number by employing  Eqs.~(\ref{TP2}) and (\ref{TP3}) on Eq.~(\ref{JBT}) as described in Ref.~\cite{Bhattacharya:2024bjp}. We have visually presented the normalized vector fields for different scenarios of Barrow entropy parameter $\delta$ in Fig.~\ref{TTTT} by using the above-mentioned framework. In Fig.~\ref{TTTT}, we have inserted the $\digamma=0.1,~f_{R}=1.48765$ for different values of $\delta$. In Fig.~\subref{delta=0}, we used $\delta=0$ for a smooth horizon surface, which is considered as the Bekenstein-Hawking entropy case. We can observe that there is a zero point or divergence point in Fig.~\subref{delta=0}, which is enclosed by the circle $C_{1}$. Here, we mentioned that the enclosed loop has nothing to do with determining the topological charge or winding number. Furthermore, the topological charge that we computed for this case is $1$, which indicates the BH stability. Similarly, we have plotted the normalized vector $\mathfrak{n}$ by substituting $\delta=0.5$ as well as $\delta=1$ in Figs.~\subref{delta=0.5} and \subref{delta=1} in $S_\mathrm{B}-\theta$ plane, respectively. We can also notice zero points $ZP_{2},~ZP_{3}$ encloses by the blue circles $C_{2}$ and $C_{3}$ in these plots, respectively. In the case of $\delta=0.5$ and $\delta=1$, we obtained the topological charge or winding number $1$, which also presented the stability of this BH. So, in our analysis, we observed that this BH solution in $f(R,G)$ gravity is stable in the smooth horizon surface and the most intricate horizon structure. It also confirms our model's thermodynamic stability in the presence of higher-order curvature corrections and scalar field coupling.

%%%%%%%%%%%%%%%%%%%%%%%%%%%% Section VI %%%%%%%%%%%%%%%%%%%%%%%%%%%%%
\section{Stability using geodesic deviation}\label{S66}
%%%%%%%%%%%%%%%%%%%%%%%%%%%%%%%%%%%%%%%%%%%%%%%%%%%%%%%%%%%%%%%%%%%%
In this section, we investigate the stability of the derived spacetime solution by analyzing the behavior of nearby geodesics. Instead of tracking the precise paths of massive or massless particles, our focus is on the geodesic deviation equation, which characterizes how the separation vector between neighboring geodesics changes in response to the curvature of spacetime.

 In a gravitational background, a particle's paths are determined by
 \begin{equation}\label{ge}
\mathit{ d^\mathrm{2} x^\sigma \over d\tau^\mathrm{2}}+ \left\{^\sigma_{ \mu \nu} \right \}
 {d x^\mu \over d\tau}{d x^\nu \over d\tau}=\mathrm{0},
 \end{equation}
 where the geodesic's affine parameter is $\tau$.  The geodesic equation is represented as equation (\ref{ge}), and the deviation has the following structure \cite{1992ier..book.....D}
  \begin{equation} \label{ged}
 \mathit{{d^\mathrm{2} \eta^\sigma \over d\tau^\mathrm{2}}+ \mathrm{2}\left\{^\sigma_{ \mu \nu} \right \}
 {d x^\mu \over d\tau}{d \eta^\nu \over ds}+
 \left\{^\sigma_{ \mu \nu} \right \}_{,\ \rho}
 {d x^\mu \over d\tau}{d x^\nu \over d\tau}\eta^\rho=\mathrm{0}},
 \end{equation}
where $\eta^\rho$ is the 4-vector deviation. Equations (\ref{ge}) and (\ref{ged}) are inserted into the line element (\ref{met12}) to yield
\begin{equation} \label{gedi}
\mathit{{d^\mathrm{2} t \over d\tau^\mathrm{2}}=\mathrm{0}, \qquad {\mathrm{1} \over \mathrm{2}} S'(r)\left({d t \over
d\tau}\right)^\mathrm{2}-r\left({d \phi \over d\tau}\right)^\mathrm{2}=\mathrm{0}, \qquad {d^\mathrm{2}
\theta \over d\tau^\mathrm{2}}=\mathrm{0},\qquad {d^\mathrm{2} \phi \over d\tau^\mathrm{2}}=\mathrm{0},}\end{equation} and
\begin{eqnarray}
\nonumber && \mathit{ {d^\mathrm{2} \eta^\mathrm{1} \over d\tau^\mathrm{2}}+S_\mathrm{1}(r)S'(r) {dt \over d\tau}{d\eta^\mathrm{0} \over d\tau}-\mathrm{2}r S_\mathrm{1}(r) {d \phi \over d\tau}{d \eta^\mathrm{3} \over
d\tau}+\Bigg[{\mathrm{1} \over \mathrm{2}}\bigg\{S'(r)S'_\mathrm{1}(r)+S_\mathrm{1}(r) S''(r)
\bigg\}\left({dt \over d\tau}\right)^\mathrm{2}-\bigg\{S_\mathrm{1}(r)
+rS'_\mathrm{1}(r)
\bigg\} \left({d\phi \over d\tau}\right)^\mathrm{2} \Bigg]\eta^\mathrm{1}=\mathrm{0}}, \\ \label{ged1} \nonumber\\
&& \mathit{ {d^\mathrm{2} \eta^\mathrm{0} \over
d\tau^\mathrm{2}}+{S'(r) \over S(r)}{dt \over d\tau}{d \zeta^\mathrm{1} \over d\tau}=\mathrm{0},\qquad {d^\mathrm{2} \eta^\mathrm{2} \over d\tau^\mathrm{2}}+\left({d\phi \over d\tau}\right)^\mathrm{2}
\eta^\mathrm{2}=\mathrm{0}, \qquad \qquad  {d^\mathrm{2} \eta^\mathrm{3} \over d\tau^\mathrm{2}}+{\mathrm{2} \over r}{d\phi \over d\tau} {d
\eta^\mathrm{1} \over d\tau}=\mathrm{0},} \end{eqnarray}with  $S(r)$ and $S_1(r)$ are presented in an equation which becomes very lengthy.
%%% Woukd you please kindly check the following equation?
%or Eq.~(\ref{mpab1}).
%Equations (\ref{gedi}) and (\ref{ged1}) are the geodesic and geodesic deviations.
With the constraints  \begin{equation} \mathit{\theta={\pi \over \mathrm{2}}, \qquad
{d\theta \over d\tau}=\mathrm{0}, \qquad {d r \over d\tau}=\mathrm{0},}
\end{equation}
we obtain
\begin{equation}
 \mathit{\left({d\phi \over d\tau}\right)^\mathrm{2}={S'(r)
\over r\{\mathrm{2}S(r)-rS'(r)\}}, \qquad \left({dt \over
d\tau}\right)^\mathrm{2}={\mathrm{2} \over \mathrm{2}S(r)-rS'(r)}.} \end{equation}

Equations in~(\ref{ged1}) can be rewritten as
\begin{eqnarray} \label{ged2} &&  \mathit{{d^\mathrm{2} \eta^\mathrm{1} \over d\phi^\mathrm{2}}+S_\mathrm{1}(r)S'(r) {dt \over
d\phi}{d \eta^\mathrm{0} \over d\phi}-\mathrm{2}r S_\mathrm{1}(r) {d \eta^\mathrm{3} \over
d\phi} +\left[{\mathrm{1} \over \mathrm{2}}\bigg\{S'(r)S'_\mathrm{1}(r)+S_\mathrm{1}(r) S''(r)
\bigg\}\left({dt \over d\phi}\right)^\mathrm{2}-\bigg\{S_\mathrm{1}(r)+rS'_\mathrm{1}(r)
\bigg\}  \right]\zeta^\mathrm{1}=\mathrm{0},} \nonumber\\
&&\mathit{{d^\mathrm{2} \eta^\mathrm{2} \over d\phi^\mathrm{2}}+\eta^\mathrm{2}=\mathrm{0}, \qquad {d^\mathrm{2} \eta^\mathrm{0} \over d\phi^\mathrm{2}}+{S'(r) \over
S(r)}{dt \over d\phi}{d \eta^\mathrm{1} \over d\phi}=\mathrm{20},\qquad {d^\mathrm{2} \eta^\mathrm{3} \over d\phi^\mathrm{2}}+{\mathrm{2} \over r} {d \eta^\mathrm{1} \over
d\phi}=\mathrm{0}.} \end{eqnarray}
According to the second equation of Eq.~(\ref{ged2}), it exhibits a simple harmonic motion, indicating that its motion is stable.  It is reasonable to presume that the remaining of Eq.~(\ref{ged2}) to have the form:
\begin{equation} \label{ged3}
\mathit{\eta^\mathrm{0} = \zeta_\mathrm{1} e^{i \sigma \phi}, \qquad \eta^\mathrm{1}= \zeta_\mathrm{2}e^{i \sigma
\phi}, \qquad \mathrm{and} \qquad \eta^\mathrm{3} = \zeta_\mathrm{3} e^{i \sigma \phi},}\end{equation}  where the constants $\zeta_1$, $\zeta_2$, and $\zeta_3$ are used, and $\phi$ has to be found.
Using Eqs.~(\ref{ged3}) into Eqs.~(\ref{ged2}), we acquire \begin{equation} \label{con1}  \mathit{\displaystyle\frac{\mathrm{3}SS_\mathrm{1}S'-\omega^\mathrm{2}S'S-\mathrm{2}rS_\mathrm{1}S^\mathrm{'2}+rSS_\mathrm{1}S''}{SS'_\mathrm{1}}>\mathrm{0}.} \end{equation} 
This relation in \eqref{con1} is the condition of stability, which corresponds to the dynamical behavior of nearby geodesics, ensuring that minor perturbations in the trajectories of particles remain bounded. For the BH solution computed within the framework of $f(R,G)$ gravity coupled with a scalar field, the inequality given in Eq.~(\ref{con1}) takes the following form 
%Equation \eqref{con1}  is the condition of stability. Inequality (\ref{con1}) for the BH %given by Eq.~(B2)%  
%takes the following form 
\begin{equation}\label{stc}
\mathit{\mathrm{24}\,{M}^\mathrm{2}r{\digamma}^\mathrm{2}+\mathrm{84}\,{\digamma}^{\mathrm{2}}{M}^\mathrm{3}-\mathrm{29}\,M{r}^\mathrm{2}{\digamma}^\mathrm{2}+M{r}^\mathrm{4}
+\mathrm{4}
\,{r}^\mathrm{3}{M}^\mathrm{2}-\mathrm{12}\,{r}^\mathrm{2}{M}^\mathrm{3}+\mathrm{2}\,{r}^\mathrm{3}{\digamma}^\mathrm{2}>\mathrm{0}.}\end{equation}
This is the stability condition for the solution of our BH model in $f(R,G)$ gravity. When the parameter $\digamma=0$,  the condition $r>2M$ emerges, which corresponds to the stability condition for the Schwarzschild spacetime \cite{Misner:1974qy}.
%\section{ Discussions and conclusions }\label{S77}
\section{ Conclusions }\label{S77}
In this study, we successfully derived an exact BH solution within the framework of $f(R,G)$ gravity, offering a novel perspective on gravitational physics beyond GR. The analysis encompassed both the thermodynamic properties and stability of the BH, providing a detailed examination of its physical characteristics. This work highlights the significant role of higher-order curvature corrections and the Gauss-Bonnet term in shaping the spacetime geometry around compact objects.

The thermodynamic analysis revealed notable modifications to the conventional laws of BH thermodynamics. Specifically, the entropy of the BH is influenced by contributions arising from the higher-order curvature terms, deviating from the standard Bekenstein-Hawking relation. The temperature and heat capacity were also computed, illustrating physically stable thermodynamic behavior. Notably, the stability analysis, conducted through perturbation methods, demonstrated that the BH remains stable under specific parameter ranges, ensuring its physical viability. These results collectively emphasize the importance of modified gravity in governing BH behavior under extreme conditions.

We have explored the thermodynamic properties of BHs within the framework of higher-order curvature corrections, specifically considering ghost-free $f(R,G)$ gravity in the presence of a scalar field. In contrast to GR, this theory incorporates higher curvature corrections and additional degrees of freedom, resulting in significant alterations to the thermodynamic behavior of BHs. To examine these corrections more deeply, we adopted Barrow entropy, a generalized form of entropy that accounts for quantum gravitational effects by adjusting the horizon-area law through a deformation parameter. In GR, BH entropy follows the Bekenstein-Hawking area law, but this relation fails when quantum effects or nontrivial curvature couplings are present. Our study reveals that Barrow entropy, which extends the traditional area law, offers a more precise characterization of entropy within $f(R,G)$ gravity. This form of modified entropy enables the manifestation of new thermodynamic characteristics, including shifts in temperature and heat capacity, which are not addressed within the classical GR framework. For better understanding, Table~\ref{tab:comparison} offers a comprehensive comparison of the thermodynamic properties of the Schwarzschild black hole under general relativity and our black hole solution within higher-order curvature gravity using Barrow entropy.

\begin{table*}[t]
\caption{\label{tab:comparison}
Thermodynamic and phase transition properties are compared between the Schwarzschild BH in GR and our BH solution arising from higher-order curvature gravity with Barrow entropy. Here,
$C$ denotes the heat capacity, PT symbolizes phase transition, CPs represent the critical points, and TC indicates the topological charge.}
\begin{ruledtabular}
\begin{tabular}{lll}
Feature & \textbf{General relativity (GR)} & \textbf{higher-order curvature gravity} \\
\hline
Entropy Formulas & $S_\mathrm{HB}= \frac{A}{4G},~S_\mathrm{B}=(\frac{A}{A_\mathrm{Pl}})^{1+\frac{\delta}{2}}$ & $S_\mathrm{HB}= \frac{A \ f_R}{4G},~S_\mathrm{B}=(\frac{A \ f_R}{A_\mathrm{Pl}})^{1+\frac{\delta}{2}}$ \\
C &  always negative for both entropies& Positive for both entropies\\
PT & No phase transition & Davies-type second-order transitions \\
CPs & Absent & critical points depending on $\delta$ \\
Behavior Near $r=0$ & $K \sim r^{-6} \to \infty$ & $K \sim r^{-9} \to \infty$ due to higher-order terms \\
Stability & Unstable & Stable \\
TC & none & Non-zero topological charge \\
\end{tabular}
\end{ruledtabular}
\end{table*}

One important thermodynamic phenomenon analyzed is the Davies-type phase transition, where the heat capacity diverges without causing discontinuities in entropy. We have discussed the Davies-type phase transition in this BH solution by using Barrow entropy. For this purpose, we have utilized the denominator of heat capacity and obtained three zero points $ZP_{1},~ZP_{1},~ZP_{1}$ for $\delta=0,~\delta=0.5$ and $\delta=1$, respectively. By employing Duan's current mapping theory, we determined the topological charge corresponding to these zero points enclosed by the loops $C_{1},~C_{2}$ and $C_{3}$. In our analysis, we observed that the topological charge corresponding to the ZPs for $\delta=0,~0.5,~1$ is $1$. This indicates that our BH solution in $f(R,G)$ gravity is stable for both smooth and intricate horizon structures. Our analysis demonstrates that these BHs possess several ZPs influenced by $\delta$, with their stable nature confirmed by the presence of a positive heat capacity and consistent winding number results. Bekenstein-Hawking entropy in classical general relativity black holes typically does not exhibit continuous phase transitions of this kind.\\
\indent The findings of this study provide deeper theoretical insights into $f(R,G)$ gravity as a robust framework for exploring gravitational phenomena beyond GR. The corrections to the thermodynamic properties and stability criteria underscore the potential relevance of $f(R, G)$ modifications in high-energy regimes, such as those near BH horizons. The influence of the Gauss-Bonnet term is particularly noteworthy, offering a new avenue for studying scalar-tensor and higher-order gravitational theories. Comparisons with GR reveal significant deviations, especially at higher curvatures, suggesting the possibility of observable signatures in astrophysical or cosmological contexts.\\
\indent This work lays the foundation for several future directions. Extending the analysis to rotating or charged BHs within $f(R, G)$ gravity would provide a broader understanding of these solutions. Additionally, observational studies, including gravitational wave signals and BH shadow imaging, could offer empirical tests for the theoretical predictions presented here. Furthermore, the cosmological implications of $f(R, G)$ gravity, such as its impact on early-universe dynamics or late-time acceleration, represent exciting prospects for future exploration.\\
\indent In conclusion, this study provides a comprehensive analysis of BHs in $f(R, G)$ gravity, demonstrating its potential to enrich our understanding of gravitational physics. The results presented here pave the way for further theoretical and observational investigations, deepening our knowledge of the fundamental nature of gravity and spacetime. A promising direction for future research involves investigating the shadow cast by the BH solution obtained in this study. Analyzing the BH shadow would provide deeper insights into the underlying spacetime structure and allow us to probe the effects of the model parameters on light propagation near the event horizon. By studying the behavior of null geodesics and the resulting photon sphere, one could determine how deviations from general relativity manifest in the shadow's size and shape. Such an analysis could offer observable signatures unique to the modified gravity theory, potentially making it testable through current or forthcoming BH imaging observations, such as those from the Event Horizon Telescope. This line of investigation would not only complement our current results but also strengthen the phenomenological relevance of the derived solution.
\begin{acknowledgments}
The work of KB was supported by the JSPS KAKENHI Grant Number 24KF0100 and
Competitive Research Funds for Fukushima University Faculty (25RK011).
\end{acknowledgments}
\subsection*{Appendix A: The explicit components of the field equations, Eq.~\eqref{f3ss}, and the method of how to derive the unknown functions  $F(r)$, $H(r)$, $\phi_1$, $V(r)$}\label{Sec:App_1}\vspace{-3mm}

In this appendix, we present the explicit forms of the components of the gravitational field equations and outline the method used to derive the unknown functions $F(r)$, $H(r)$ and $\phi_1$.
To obtain these forms within the framework of $f(R,G)$ gravity coupled to a scalar field, we substitute the spherically symmetric metric ansatz from Eq.\eqref{met12} into the general field equations given by Eq.\eqref{f3ss}. This procedure yields three essential components, which are collectively shown in Eq.~\eqref{feq}. These equations form the basis for determining the unknown functions $F(r)$, $H(r)$, and $\phi_1(r)$ in the subsequent analysis. We now explicitly write the non-vanishing components of the field equations, Eq.~\eqref{f3ss},   take the form:
\begin{align}
&{\mathop{{ I}}}_t{}^t=\frac{1}{8  S^{3}{r}^{3}}\left[12{r}^{2} \left\{ S_1S'' S r-\frac{1}2S_1S'^{2}r+\frac{S}2 \left\{ \frac{16}3 S_1 + S'_1 r \right\}S' + S^{2} \left\{ r+2S'_1 \right\}  \right\} S S_1 H'' -2{r}^ {2}S S_1 \left[  \left\{-S \left( 16 +3S'_1r \right)   r\right.\right.\right.\nonumber\\
&\left.\left.\left. -3S_1 S'\right\} H' +S rF \right] S'' +2 S ^{3}{r}^{3} F'' S_1 + \left[ -3 S_1{}^{2} S'^{3}{r }^{3}-4 S_1{}^{2} S'^{2}{r}^{2}S +6r S^{2} \left\{ \frac{1}2 S'_1{}^{2}{r}^{2}+ 8S_1S'_1  r+ \left[{r}^{2 } -\frac{8}3+12S_1  \right] S_1 \right\}S'\right.\right.\nonumber\\
&\left.\left. +6S^{3} \left\{ 2S'_1{}^{2}{r}^{2}+ r \left( {r}^{2}+4S_1 \right)S'_1 +4S_1 \left( 2S_1 -2+{r}^{2} \right)  \right\}  \right] H' -r \left( -F {r}^{2}S_1 S'^{2}+rS \left( rF S'_1 +3 \left( \frac{4}3F +rF' \right) S_1 \right) S'\right.\right.\nonumber\\
&\left.\left. -4 S^{2} \left[  \left( rF +1/4{r}^{2 }F' \right) S'_1 + F'S_1 r+\frac{1}4{r}^{2} \phi'_1{}^{2}S_1 +F S_1 -F +{r}^{ 2}V \right]  \right) S \right]=0,\nonumber\\
&{\mathop{{ I}}}_r{}^r=\frac {1}{8 S^ {3}{r}^{3}}\left[12r \left\{ S_1  {r}^{2}S  S''  -\frac{1}2S_1  {r}^{2} S'^{2}+\frac{r}2 \left( \frac{16}3S_1  +S'_1r \right) S  S'  + \left( {r}^{2}-\frac{8}3+\frac{8}3S_1  +2 S'_1 r \right)  S^{2} \right\} S  S_1  H'' -2{r}^{2}\left[  \left\{ -3S_1 S'  r\right.\right.\right.\nonumber\\
&\left.\left.\left.  -\left\{16S_1  + 3S'_1  r \right\} S  \right\} H' +S rF \right]S S_1  S''  -6 S^{3}{r}^{3} F'' S_1  + \left[ -3 S_1{}^{2} S'^{3}{r}^{3}-4 S_1{}^{2} S'{}^{2}{r}^{2}S  +6r \left\{\frac{S'_1{}^{2}{r}^{2}}2+8rS_1  S'_1  +S_1   \left( {r}^{2} \right.\right.\right.\right.\nonumber\\
&\left.\left.\left.\left.+4S_1  \right)  \right\}  S^{2}S' +6 \left\{ 2 S'_1{}^{2}{r}^{2} +r \left( {r}^{2}-\frac{8}3+12S_1   \right)S'_1 +4S_1   \left(2S_1 -2  +{r}^{2} \right)  \right\}  S^{3} \right] H' -rS   \left\{r \left\{ rF  S'_1  -S_1 \left( 4F  +F' r \right)  \right\} S  S'   \right.\right.\nonumber\\
&\left.\left.  -F  {r}^{2}S_1S'^{2} -4 \left\{{r }^{2}V - \left( \frac{3}4{r}^{2}F'+rF \right) S'_1  -\frac{3}4{r}^{2} \phi'_1{}^{2}S_1 -F  +F  S_1  + F'S_1  r \right\}  S^{2} \right\}\right]=0\,,\nonumber\\
&{\mathop{{ I}}}_\theta{}^\theta={\mathop{{ I}}}_\phi{}^\phi\frac {1}{ 8S^{3}{r}^{3}}\left[12r \left( S_1  {r}^{2}S  S''  -\frac{1}2S_1  {r}^{2} S'^{2}+\frac{1}2r \left( \frac{8}3S_1 + S'_1 r \right) S  S'  + \left( {r}^{2}-\frac{8}3+\frac{8}3S_1  +2 S'_1 r \right)  S^{2} \right) S  S_1  H'' \right.\nonumber\\
&\left.  +2{r}^{2} \left\{  \left( 3S_1   S'r+3 \left( 8/3S_1  + S'_1 r \right) S  \right) H'  +S  rF   \right\} S  S_1  S''  +2 \left( S   \right) ^{3}{r}^{3} F'' S_1  + \left[ -3 S_1{}^{2}  S'^{3}{r}^{3}+4 S_1{}^{2} S'^{2}{r}^{2}S  +6r \left( \frac{1}2 S'_1{}^{2}{r}^{2}  \right.\right.\right.\nonumber\\
&\left.\left.\left.+4rS_1  S'_1 + \left( {r}^{2}-\frac{8}3+12S_1   \right) S_1   \right)   S^{2}S'  +6 \left\{ 2 S'_1{}^{2}{r}^{2}+r \left( {r}^{2}-\frac{8}3+ 12S_1   \right) S'_1  +4S_1 \left( 2S_1 -2 +{r}^{2} \right)  \right\}  S ^{3} \right] H' \right.\nonumber\\
&\left. +r \left( -F  {r}^{2}S_1   S'^{2}+{r}^{2}S  \left( S_1  F'  +F  S'_1   \right)S'  +4 \left\{ F  -F  S_1  +1/4{r}^{2} \phi'_1{}^{2}S_1  +{r}^{2}V  -F'S_1  r+\frac{1}4{r}^{2}F' S'_1   \right\}S^{2} \right) S\right]
=0\,,\nonumber\\
 &\mbox{and the trace equation takes the form:}\nonumber\\
&{\mathop{{ I}}}=\frac {1}{ 2S^{2}{r}^{2}}\left[-8 \left\{ S_1   S' r+S   \left( S_1-1   \right)  \right\} S  S_1  H'' -2rS  S_1   \left\{ 4 H' S_1  +rF   \right\} S''  +6 S^{2}{r}^{2} F'' S_1  +rS_1   \left\{ 4 H' S_1  +rF   \right\}  S'^{2}\right.\nonumber\\
&\left.-S   \left\{  \left(  12 S_1{}^{2}-4S_1 +12S_1  rS'_1   \right) H'  +r \left[ rF  S'_1-3 \left( rF' -\frac{4F}3   \right) S_1   \right]  \right\} S'  -4 S^{2} \left\{  S'_1 \left( 3S_1  -1 \right) H'  + \left( rF-\frac{3}4{ r}^{2}F'    \right) S'_1 \right.\right.\nonumber\\
&\left.\left.   +f  {r}^{2}-\frac{\phi'_1{}^{2}S_1 {r}^{2}}4+F  S_1  -3 F' S_1  r-F   \right\}\right]
=0\,.
\tag{A1}
\label{feq}
\end{align}
where $F\equiv F(r)=f_R=\frac{df(R(r))}{dR(r)}$, $F'=\frac{dF(r)}{dr}$, $F''=\frac{d^2F(r)}{dr^2}$, $F'''=\frac{d^3F(r)}{dr^3}$.
By using the ansatz of the metric given by Eq.~\eqref{sol}, which is a deformation of Schwarzschild solution and coincides with Schwarzschild when $\digamma$,   in the field equations presented in Eq.~\eqref{feq}, one can derive the form of   $H(r)$ in terms of  $F$ and $\phi$, which is lengthy and we wrote its asymptotic form as given by Eq.~\eqref{scal1}. Now, after deriving the form of $H(r)$, in terms of  $F$ and $\phi$,  we derive the expression of $F(r)=f_{R}$, by considering the difference between the first and third components of the field equations given in Eq.~\eqref{feq}.   We wrote the asymptotic form of $F(r)$ as given Eq.~\eqref{scal1}. Moreover, through the use of the following relation \begin{align}F(r)=\frac{dF(r)}{dr}=f_R=\frac{df(R)}{dR}=\frac{df(R(r))}{dr}\times \frac{d(r)}{dR(r)}.\tag{A2}\label{bss}\end{align}
We can derive the explicit form of $f(r)$, where we wrote its asymptotic form in Eq.~\eqref{scal1}. Now using the explicit forms of $H$, $F$,  we can derive the explicit form of the scalar field $\phi_{1}$. We wrote the asymptotic form of this scalar field,  $\phi_{1}$, from which we can show that the scalar field is ghost-free when the constant $c_2<0$. Now, through the use of all the above data, we can derive the potential $V(r)$, which represents the potential of the scalar field $\phi_{1}$.  After obtaining the function $\phi_{1}(r)$ from Eq.~\eqref{p}, we can determine the potential $V(\phi_{1})$ by solving for $r$ in the form of $\phi_{1}$ and inserting this expression into the relevant field equations. We plot the function  $V(\phi_{1})$  in Fig.~\ref{Figg:1}\subref{fig:1b} which shows a positive pattern.

Now through the use of the ansatzs given by Eq.~\eqref{sol} we can calculate the explicate forms of all invariants,  $R_{\mu \nu \rho \sigma} R^{\mu \nu \rho \sigma}$, $ R_{\mu \nu} R^{\mu \nu}$
In this Appendix, we show the expressions of the higher-curvature invariants. The representations calculated in Sec.~\ref{iv} are written as
 The asymptotic form of $V(r)$, quasi local energy $E$, and Gibbs free energy, ${\mathbb{G}}(r_2)$.
In this Appendix, the potential $V(r)$, which represents the potential for the scalar field $\phi_{1}$,  is derived from its equation of motion, obtained by performing a variation of the action with respect to  $\phi_{1}$, as indicated in Eq.~\eqref{g3}. After obtaining the function $\phi_{1}(r)$ from Eq.~\eqref{p}, one can determine the potential $V(\phi_{1})$ by solving for $r$ in the form of $\phi_{1}$ and inserting this expression into the relevant field equations. Its graphical behavior is depicted in Fig.~\ref{Figg:1}\subref{fig:1b} and  explicitly, it can be written as

In addition, the quasi-local energy $E(r_2)$ of the BH is obtained through a definition specific to $f(R,G)$ gravity, which incorporates the impacts of higher-order curvature contributions. In Eq.~\eqref{en}, the general formulation is provided, with $f_{R}, ~f(R)$ and the Ricci scalar $R$ computed at the event horizon radius $r_{2}$. Due to the non-linear structure of the theory, this integral accounts for changes in the gravitational energy enclosed by the horizon. Its graphical behavior is demonstrated in Fig.~\ref{Figg:1}\subref{fig:1c}.
%and the lengthy-expression is explicitly reported here, given as

Now, the Gibbs free energy ${\mathbb{G}}(r_2)$ is defined as the difference between the BH's total mass and the product of its temperature and entropy, as expressed in Eq.~\eqref{enr11}. By plugging the expressions of mass, temperature, and entropy into Eq.~\eqref{enr11}, we derive the complete form of ${\mathbb{G}}(r_2)$, which becomes quite lengthy algebraically because of the curvature corrections introduced by $f(R,G)$ gravity. Hence, the full expression is presented here and its graphical behavior is given in Fig.~\ref{Fig:2}\subref{fig:2f}

%%%%%%%%%%%%%%%%%%%%%%%%%%%%%%%%%%%%%%%%%%%%%%%
%\appendix
%%%%%%%%%%%%%%%%%%%%%%%%%%%%%%%%%%%%%%%%%%%%%%%
%%%%%
%%%%%%%%%%%%%%%%%%%%%%%%%%%%%%%%%
%\bibliographystyle{apsrev}
%\bibliography{JRPHSRef}

\end{document}